\documentclass[aps,twocolumn,amsmath,amssymb,preprintnumbers,floatfix,prb,superscriptaddress,longbibliography]{revtex4-2}

\usepackage{comment}
\usepackage[version=4]{mhchem}
\usepackage[utf8]{inputenc}
\usepackage{newtxtext}
\usepackage[upint]{newtxmath}
\usepackage{microtype}
\usepackage{textcomp}
\usepackage{dsfont}
\usepackage{eucal}
\usepackage{siunitx}
\usepackage{soul}

\usepackage{enumerate}
\usepackage{amsfonts}
\usepackage{color}
\usepackage{soul}
\usepackage{cancel}

\usepackage{todonotes}
\presetkeys%
    {todonotes}%
    {inline}{}

\usepackage{graphicx}

\usepackage[colorlinks,allcolors=blue]{hyperref}
\usepackage[capitalize]{cleveref}

\makeatletter
\providecommand{\href@noop}[2]{#2}
\makeatother

\definecolor{DarkBlue}{rgb}{0,0,0.80}
\definecolor{DarkRed}{rgb}{0.80,0,0}
\definecolor{Purple}{rgb}{0.55,0,0.55}
\definecolor{Purple}{rgb}{0,0,0.8}

\newcommand{\Chi}[1]{\textcolor{DarkBlue}{#1}}

\newcommand{\Chicomment}[1]{\textcolor{DarkBlue}{\textbf{[COMMENT: #1]}}}

\newcommand{\jacob}[1]{\textcolor{DarkRed}{#1}}     
\newcommand{\jcomment}[1]{\textcolor{DarkRed}{\textbf{[COMMENT: #1]}}}

\DeclareMathOperator{\im}{Im}

\newcommand{\B}[1]{\bm{#1}}

\let\epsilon\varepsilon
\begin{document}
\title{
Turning Zeeman splitting into switchable charge polarization in a double quantum dot}
\author{Chi Sun}
\affiliation{Aix-Marseille Université, CNRS, CINaM, Marseille, France}
\affiliation{Center for Quantum Spintronics, Department of Physics, Norwegian \\ University of Science and Technology, NO-7491 Trondheim, Norway}
\author{Jacob Linder}
\affiliation{Center for Quantum Spintronics, Department of Physics, Norwegian \\ University of Science and Technology, NO-7491 Trondheim, Norway}
\begin{abstract}
A magnetic field that acts identically on two quantum dots is not expected to move charge between them. Nevertheless, we show that a uniform Zeeman field can strongly reconfigure and even reverse the single-electron charge polarization of an asymmetric open double quantum dot. Using a symmetry-preserving Green's-function equation-of-motion approach, we identify regimes where the preferred dot occupation reverses while the system remains in the single-electron charge sector. Two distinct mechanisms produce this behavior. Unequal gate levels produce different occupation responses because the Zeeman-shifted resonances lie at different positions relative to the reservoir chemical potential, whereas unequal onsite interactions distinguish the dots through their many-body addition spectra. Coulomb blockade stabilizes the single-electron sector, causing the reservoir-mediated response to appear as spatial charge redistribution rather than a change in total occupation. Our results establish a mechanism for magnetic control of charge polarization at fixed electrostatic detuning.

\end{abstract}

\maketitle

\textit{Introduction---}Quantum dots (QDs) are nanoscale structures that confine electrons in all
spatial dimensions, producing discrete electronic states analogous to
those of atoms \cite{QD_research,reimann2002electronic}. Their controllable charge and spin degrees of
freedom provide a versatile platform for exploring few-electron physics
and quantum transport \cite{loss1998quantum,ledentsov1998quantum,garcia2021semiconductor}. Coupling two QDs forms a double quantum dot (DQD),
introducing an additional spatial degree of freedom where interdot
coupling and Coulomb interactions give rise to rich charge and spin
phenomena. The resulting interplay of charge,
spin, and interactions makes DQDs a paradigmatic system for studying
correlated quantum dynamics and quantum computation \cite{sobrinoprb24,burkard1999coupled,DQD_science,van2002electron,DQD_qubit,PRXQuantum.6.020329}.\\

Charge and spin control in DQDs has been pursued through distinct routes.
Charge states are typically manipulated electrically through detuning,
tunnel coupling, or microwave excitation \cite{microwave_1,PhysRevLett.91.226804}, whereas magnetic
fields have primarily been used to control spin and exchange interactions \cite{burkard1999coupled,stopa2008magnetic,PhysRevB.73.045313}.
Magnetic-field-induced charge redistribution within a single QD 
has also been reported in
disordered graphene \cite{graphene}, arising from an orbital magnetic response. A spatially uniform Zeeman splitting presents a more fundamental puzzle. It shifts the spin levels of both dots identically and is therefore proportional to the identity matrix in the left-right orbital subspace. Within an isolated single-electron picture, such a field can polarize the spin but has no means of deciding whether the electron should occupy the left or the right dot. Magnetic control of the left-right charge polarization in DQDs would therefore seem to be ruled out.\\

In this work, we show that this expectation is overturned in an open interacting DQD. Using
a symmetry-preserving Green's-function equation-of-motion (EOM) approach,
we show that a uniform Zeeman field can strongly redistribute the
single-electron occupation between the two dots while the DQD remains
predominantly in the single-electron regime. For unequal gate levels,
the field produces a pronounced modulation of the left-right charge
polarization. More remarkably, unequal onsite Coulomb interactions enable
a complete reversal of the preferred dot occupation even when the bare
gate levels are nearly identical. All of this occurs despite the Zeeman field being applied uniformly to the DQD. This reveals that interactions can
convert a uniform spin splitting, which carries no intrinsic
left-right bias, into directional charge switching. Our results establish
Zeeman-driven charge switching as a distinct manifestation of coupled
spin and charge dynamics in interacting DQDs, providing an unexpected way of controlling spatially resolved charge occupation in DQDs without changing the electrostatic detuning.\\


\begin{figure}
\includegraphics[width=0.85\columnwidth]{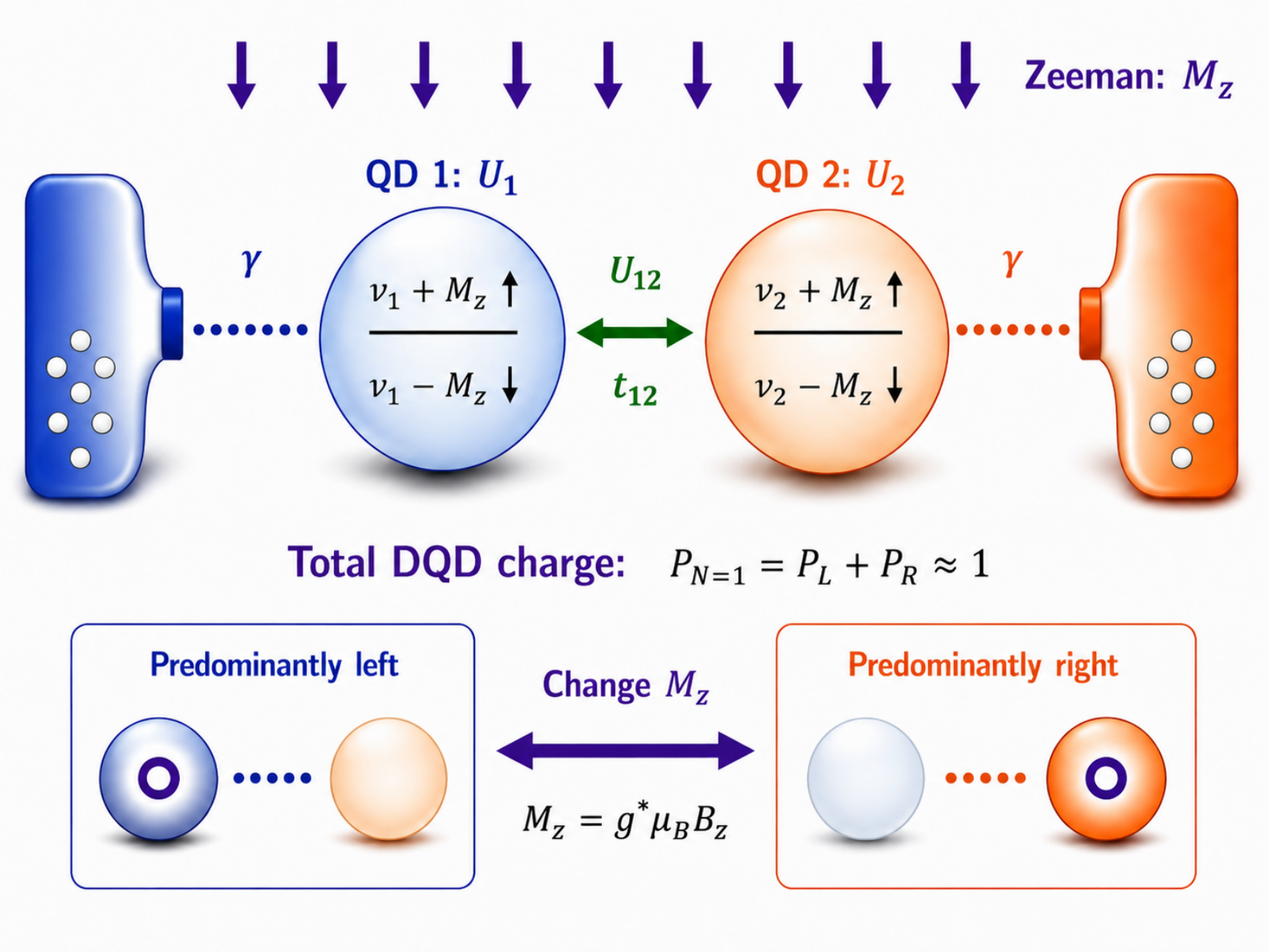}
	 \caption{Double quantum dot (DQD) charge-qubit setup.
Two interacting QDs, with onsite Coulomb interactions $U_{1,2}$, are coupled by the interdot interaction $U_{12}$ and tunneling $t_{12}$, and separately coupled to electronic reservoirs with strength $\gamma$. A uniform Zeeman field $M_z$ shifts the gate levels as $v_{i\sigma}=v_i+\sigma M_z$. In the single-electron regime, $P_{N=1}=P_L+P_R\simeq1$, varying $M_z$ controls the charge distribution between the two QDs.}
    \label{fig:model}
\end{figure}

\textit{Theory---}
We consider an interacting DQD coupled to two
reservoirs (see Fig. \ref{fig:model}) with $\hat H =\hat H_{\rm leads}+\hat H_{\rm 2dot}+\hat H_{\rm tun}$:
\begin{align}
\hat H_{\rm leads}
&=
\sum_{ik\sigma}
\varepsilon_{ik\sigma}
\hat c_{ik\sigma}^{\dagger}\hat c_{ik\sigma}, \notag\\
\hat H_{\rm tun}
&=
\sum_{ik\sigma}
\left(
V_{ik\sigma}\hat c_{ik\sigma}^{\dagger}\hat d_{i\sigma}
+{\rm H.c.}
\right),\notag\\
\hat H_{\rm 2dot}
&=
\sum_{i\sigma}v_{i\sigma}\hat n_{i\sigma}
+\sum_i U_i\hat n_{i\uparrow}\hat n_{i\downarrow}
+U_{12}\hat n_1\hat n_2
\notag\\&+t_{12}\sum_{\sigma}
\left(
\hat d_{1\sigma}^{\dagger}\hat d_{2\sigma}
+{\rm H.c.}
\right).
\end{align}
Here, $\hat d_{i\sigma}$ ($\hat c_{ik\sigma}$) annihilates an electron
with spin $\sigma$ in QD $i$ (the corresponding reservoir),
$\hat n_{i\sigma}=\hat d_{i\sigma}^{\dagger}\hat d_{i\sigma}$, and
$\hat n_i=\sum_{\sigma}\hat n_{i\sigma}$. The parameters $U_i$ and
$U_{12}$ denote the intra- and interdot Coulomb repulsions, respectively,
while $t_{12}$ is the interdot tunneling amplitude. The reservoir states
have energies $\varepsilon_{ik\sigma}$, and $V_{ik\sigma}$ denotes the
tunneling between QD $i$ and its corresponding reservoir. A spatially uniform magnetic field
$\boldsymbol B=(0,0,B_z)$ enters through $v_{i\sigma}=v_i+\sigma M_z$ with $M_z=g^{*}\mu_BB_z$,
where $v_i$ represents the dot level (gate energy). Importantly, the same
Zeeman splitting $M_z$ acts on both dots and therefore does not itself
introduce a left-right energy bias.\\

To resolve the correlated many-body occupations, we employ a
Green's-function EOM approach \cite{EOM,EOMsym}. 
The retarded single-particle Green's function (GF) and its
Fourier transform are defined as
\begin{equation}
G^r_{i\sigma}(t)
=
-i\Theta(t)
\langle
\{\hat d_{i\sigma}(t),\hat d_{i\sigma}^{\dagger}(0)\}
\rangle ,
\end{equation}
\begin{equation}
G^r_{i\sigma}(\omega)
\equiv
\langle\!\langle
\hat d_{i\sigma}:\hat d_{i\sigma}^{\dagger}
\rangle\!\rangle_\omega
=
\int_{-\infty}^{+\infty}
G^r_{i\sigma}(t)e^{i\omega t}dt .
\end{equation}
For a general operator $\hat B$, the corresponding EOM reads
\begin{equation}
\hbar\omega_+
\langle\!\langle
\hat B:\hat d_{i\sigma}^{\dagger}
\rangle\!\rangle_\omega
=
\hbar\langle
\{\hat B,\hat d_{i\sigma}^{\dagger}\}
\rangle
+
\langle\!\langle
[\hat B,\hat H]:\hat d_{i\sigma}^{\dagger}
\rangle\!\rangle_\omega ,
\end{equation}
where $\omega_+=\omega+i\eta$ with $\eta\rightarrow0^+$ \cite{SM}. 
Taking $\hat B=\hat D\hat d_{i\sigma}$ and applying the EOM
successively generates higher-order GFs
$\langle\!\langle\hat D\hat d_{i\sigma}:
\hat d_{i\sigma}^{\dagger}\rangle\!\rangle_\omega$, where
$\hat D=\hat n_{i_1\sigma_1}\hat n_{i_2\sigma_2}\cdots
\hat n_{i_N\sigma_N}$ is a product of $N$ density operators with $i_N\sigma_N\neq i\sigma$ since $\hat{n}_{i\sigma}\hat{d}_{i\sigma}=0$. We focus on the interaction-dominated regime, where the dot-reservoir
coupling and interdot tunneling are smaller energy scales than the Coulomb
interactions. We close the EOM in the occupation-diagonal density sector, appropriate
to the interaction-dominated regime considered here. The approximation
retains the direct effect of finite interdot tunneling on the
density-sector propagation while truncating the additional
tunneling-induced hierarchy of interdot coherences. Details and
extensive consistency checks are provided in the Supplemental Material \cite{SM}. The interacting problem is thereby mapped onto a truncated hierarchy of
density-dressed GFs involving one-, two-, and three-body density
correlators. The coupling to the reservoirs enters this hierarchy through
the hybridization function $\Delta_{i\sigma}(\omega)
=
\sum_k
\frac{|V_{ik\sigma}|^2}
{\hbar\omega_+-\varepsilon_{ik\sigma}}$,
which reduces to $\Delta_{i\sigma}=-i\gamma/2$ in the wide-band limit. Unlike mean-field theory which only covers the first-order density correlator, the EOM approach retains independent higher-order density correlations and the resulting interaction-dependent many-body resonances. Further details are given in SM \cite{SM}.\\

The next step relates the density correlators back to the corresponding
GFs. The spectral relation
gives
\begin{equation}
\langle\hat D\hat n_{i\sigma}\rangle
=
-\int\frac{d\omega}{\pi}\,
f(\omega)\,
{\rm Im}
\left[
\langle\!\langle
\hat D\hat d_{i\sigma}:
\hat d_{i\sigma}^{\dagger}
\rangle\!\rangle_\omega
\right],
\end{equation}
in which $f(\omega)$ is the Fermi distribution. Thus, the EOM hierarchy expresses the GFs in terms of
density correlators, while the spectral relation determines these
correlators from the GFs. Combining the two closes the problem
self-consistently in terms of density correlators. As a result, the one-, two-, and
three-body correlators form a $14\times14$ linear system $\boldsymbol M \boldsymbol y= \boldsymbol b,$ 
where $\boldsymbol y$ contains four one-body, six two-body, and four
three-body density correlators. The explicit forms of $\boldsymbol y$,
$\boldsymbol M$, and $\boldsymbol b$ are given in the Supplemental Material, in which a symmetry-preserving treatment \cite{EOMsym} is applied to remove the artificial truncation-induced asymmetry. Given the solution of $\boldsymbol y$ by solving the above equation, the four-body correlator $\langle
\hat n_{1\uparrow}\hat n_{1\downarrow}
\hat n_{2\uparrow}\hat n_{2\downarrow}
\rangle $ can be obtained from the highest-order EOM in terms of the three-body
correlators \cite{SM}. Together, these equations close the density-correlation hierarchy for the four occupation modes.\\

This complete set of one- through four-body density correlations allows us to
reconstruct the probability of every many-body occupation state
$|ABCD\rangle$, where $A,B,C,D\in\{0,1\}$ denote the occupations of
$(1\uparrow,1\downarrow,2\uparrow,2\downarrow)$ \cite{SM}:
\begin{align}
P_{ABCD}
&=
\langle
\hat n_{1\uparrow}^{A}(1-\hat n_{1\uparrow})^{1-A}
\hat n_{1\downarrow}^{B}(1-\hat n_{1\downarrow})^{1-B}\notag\\
&\hat n_{2\uparrow}^{C}(1-\hat n_{2\uparrow})^{1-C}
\hat n_{2\downarrow}^{D}(1-\hat n_{2\downarrow})^{1-D}
\rangle .
\end{align}
Our formulation therefore resolves the full 16-state correlated
occupation distribution \cite{SM}, rather than only the average occupations
$\langle\hat n_{i\sigma}\rangle$ obtained in standard mean-field {GF} 
approaches. In particular, within the single-electron 
subspace, we define the left- and right-localized probabilities as $P_L=P_{1000}+P_{0100}$ and $P_R=P_{0010}+P_{0001}$ with $P_{N=1}=P_L+P_R$ being
the total probability of occupying the single-electron ($N=1$) sector. \\

The full occupation distribution above allows us to distinguish changes in the total DQD charge from redistribution within the single-electron sector. We therefore ask whether a spatially uniform Zeeman field can control the left-right charge polarization. At the single-particle level, this appears unlikely because the same Zeeman shift acts on both dots and leaves their bare detuning unchanged. In an open interacting DQD, however, the field modifies the spin-resolved addition and removal processes involving the reservoirs. If the dots differ in their gate level positions or Coulomb energies, these processes need not affect them equally. Coulomb blockade can then convert the unequal response primarily into a redistribution of single-electron weight within the \(N=1\) sector. As shown in detail below, this mechanism produces continuous magnetic charge reconfiguration and, in suitable regimes, even reverses the preferred dot occupation.\\

\begin{figure}
\includegraphics[width=\columnwidth]{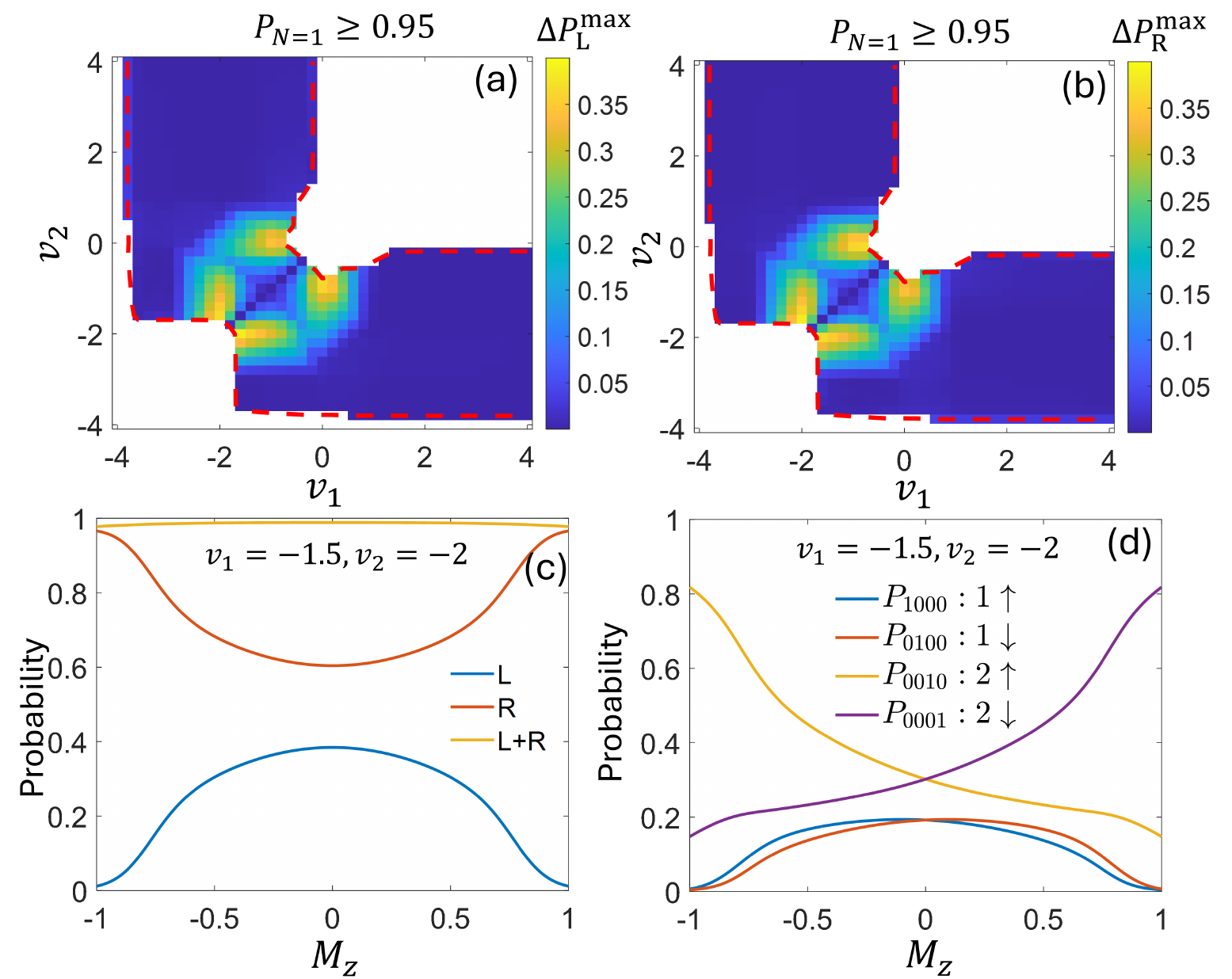}
	\caption{Magnetic charge reconfiguration for $U_1=U_2$.
(a,b) Maximum field-induced changes
$\Delta P_{L,R}^{\max}(v_1,v_2)
=\max_{|M_z|\leq1}|P_{L,R}(v_1,v_2;M_z)-P_{L,R}(v_1,v_2;0)|$,
restricted to $P_{N=1}\geq0.95$ throughout the field sweep $|M_z|\leq 1$.
(c) Left, right, and total single-electron probabilities and
(d) the four spin-resolved single-electron probabilities versus $M_z$
at $v_1=-1.5$ and $v_2=-2$.
Parameters are $U_1=U_2=4$, $U_{12}=3$, $t_{12}=0.3$,
$T=0.05$, and $\gamma=0.05$, with $\Delta_{i\sigma}=-i\gamma/2$.
The red dashed line in (a,b) denotes the boundary $P_{N=1}=0.95$ and the white regions outside the red dashed line do not satisfy the $P_{N=1}\geq0.95$ criterion.} 
    \label{fig:reconfig}
\end{figure}

\textit{Magnetic reconfiguration at equal onsite Coulomb interactions} --- We first consider equal onsite Coulomb interactions, \(U_1=U_2\), while retaining a gate level asymmetry, $v_1\neq v_2$. This choice isolates the role of the unequal dot levels from that of interaction asymmetry. Figures \ref{fig:reconfig}(a) and \ref{fig:reconfig}(b) map the maximum field-induced changes
$\Delta P_{L}^{\max}(v_1,v_2)=\max_{|M_z|\leq1}|P_L(v_1,v_2;M_z)-P_L(v_1,v_2;0)|$
and $\Delta P_R^{\max}$, respectively. Pronounced responses within the $N=1$ sector extend over broad regions of the $(v_1,v_2)$ plane, demonstrating that the effect does not require a fine-tuned operating point. At the representative point $v_1=-1.5$ and $v_2=-2$, Fig. \ref{fig:reconfig}(c) shows a substantial transfer of occupation between the two dots while retaining $P_L+P_R\simeq1$. Here $P_R>P_L$ throughout the field sweep $|M_z|\leq1$, so the field strongly modulates the left-right charge polarization without reversing its sign. This charge redistribution originates from a reorganization of the spin-resolved many-body configurations [Fig. \ref{fig:reconfig}(d)]: the dominant right-dot state evolves from $P_{0010}$ for negative $M_z$ to $P_{0001}$ for positive $M_z$, with the spin partners becoming degenerate at $M_z=0$. Thus, a spatially uniform Zeeman splitting can be converted into a sizable left-right charge redistribution in a level-asymmetric open DQD despite containing no intrinsic left-right asymmetry. In contrast, for $v_1=v_2$ the two dots respond identically and the charge redistribution vanishes, producing the darkest-blue diagonal ($\Delta P_L^\text{max}=\Delta P_R^\text{max}=0$) in Figs. \ref{fig:reconfig}(a) and \ref{fig:reconfig}(b).\\ 

The origin of the charge redistribution can be understood by first considering the noninteracting limit $U_1=U_2=U_{12}=0$. In an open DQD, the spin-summed charge occupation of dot \(i\) can be written schematically as 
\begin{equation} 
\langle\hat{n}_i(M_z)\rangle \simeq F(v_i-M_z)+F(v_i+M_z), 
\end{equation} 
where $F(E)$ describes the occupation of a dot level, including both the Fermi--Dirac distribution of the reservoir and the broadening caused by dot--reservoir coupling. Because $F(E)$ varies nonlinearly near the reservoir chemical potential, the Zeeman-shifted levels of two dots with \(v_1\neq v_2\) can undergo different changes in occupation through the many-body interaction, even though \(M_z\) acts equally on both dots and leaves their bare detuning unchanged. This is particularly transparent when one dot level lies close to the Fermi edge while the other lies farther away: the spin-resolved resonances of the near-resonant dot are then strongly affected by \(M_z\), whereas those of the other dot respond more weakly. In the absence of Coulomb blockade, this mechanism would predominantly manifest itself as a change in the total DQD occupation, for example through an \(N=0\) to \(N=1\) loading transition. In our case, Coulomb interactions qualitatively change the resulting response by energetically suppressing configurations with additional electrons ($N>1$) and thereby stabilizing the single-electron sector,  $P_{N=1}=P_L+P_R\simeq1.$ Once the total DQD occupation is constrained in this way, the unequal Zeeman-dependent loading and removal processes of the two dots induced by $v_1\neq v_2$ are expressed primarily as a redistribution between the left- and right-localized single-electron configurations $P_L$ and $P_R$, rather than as a large change in the total charge $0\leq N\leq4$. The reservoirs therefore provide the exchange processes required to reorganize the single-electron occupations, while the Coulomb interactions prevent these processes from producing a substantial change in the total DQD charge. Further, interactions also introduce occupation-dependent resonances associated with electron addition and removal. Their energies and spectral weights depend on the spin-resolved occupations and higher-order density correlators, which are fully accessible within our EOM treatment. These additional interacting channels can modify the unequal response of the two dots due to $v_1\neq v_2$ and may contribute to its magnitude. The observed behavior can therefore be interpreted as an open-system conversion of a spatially uniform spin splitting into left-right charge polarization. As for the fully symmetric limit, \(v_1=v_2\) and otherwise identical dot and reservoir parameters, the two Zeeman responses are identical and the redistribution vanishes [see the diagonal darkest blue regions in Fig. \ref{fig:reconfig}(a,b)], as required by left-right symmetry.\\

\begin{figure}
\includegraphics[width=\columnwidth]{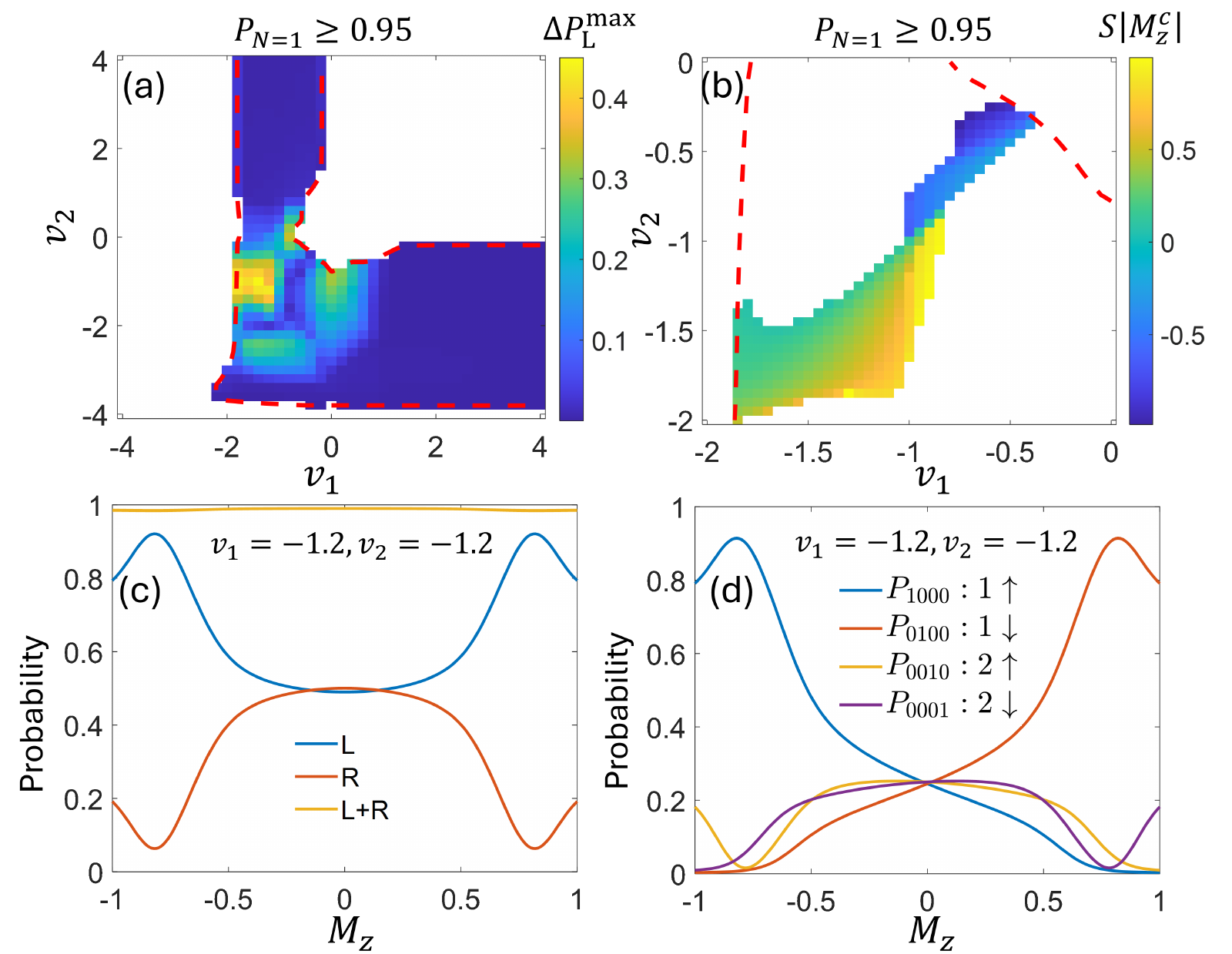}
	\caption{Magnetic charge switching for $U_1\neq U_2$.
(a) Maximum charge redistribution $\Delta P_L^{\max}$.
(b) Signed switching threshold $S|M_z^c|$, where
$P_L(M_z^c)=P_R(M_z^c)$ and $S>0$ ($S<0$) denotes right-to-left
(left-to-right) switching.
(c) Left, right, and total single-electron probabilities and
(d) spin-resolved probabilities versus $M_z$ at $v_1=v_2=-1.2$.
The red dashed line in (a,b) denotes the boundary $P_{N=1}=0.95$ throughout $|M_z|\leq1$ and the white regions outside the red dashed line do not satisfy the $P_{N=1}\geq0.95$ criterion.
Here $U_1=2$, $U_2=4$, and $U_{12}=3.5$; other parameters are as in
Fig. \ref{fig:reconfig}.} 
    \label{fig:switch}
\end{figure}

\textit{Magnetic switching at unequal onsite Coulomb interactions ---} We next consider nearly equal bare dot levels, \(v_1\simeq v_2\), but unequal onsite interactions, \(U_1\neq U_2\). This case is qualitatively distinct from the level-asymmetric situation in Fig. \ref{fig:reconfig}. Figures \ref{fig:switch} and \ref{fig:switch}(b) show,
respectively, the maximum field-induced redistribution $\Delta P_L^{\max}$
and the threshold of the Zeeman field causing switching $S|M_z^c|$, where
$P_L(M_z^c)=P_R(M_z^c)$. Here $S>0$ ($S<0$) denotes right-to-left
(left-to-right) switching by adding a nonzero $M_z$. A pronounced switching region emerges around
$v_1\simeq v_2$, where the two dots are nearly indistinguishable at the
bare one-electron level. Moreover, the sign change of $S|M_z^c|$ across
the $(v_1,v_2)$ plane shows that the gate levels control not only the
switching threshold but also its direction. The emergence of switching
near $v_1=v_2$ therefore points to a mechanism beyond bare level
asymmetry as discussed in the previous section.\\

In an isolated DQD constrained strictly to the \(N=1\) sector, double occupation is absent, so that ${\langle}\hat n_{i\uparrow}\hat n_{i\downarrow}{\rangle}=0$, while ${\langle}\hat n_1\hat n_2{\rangle}=0$ because a single electron cannot occupy both dots simultaneously. The projected one-electron Hamiltonian is therefore independent of \(U_1\), \(U_2\), and \(U_{12}\). For $v_1=v_2$ and a spatially uniform $M_z$, the spin and orbital parts of the Hamiltonian separate, and $M_z$ cannot modify the left-right charge distribution. Consequently, \(P_L=P_R\) holds in the strictly isolated and left-right-symmetric \(N=1\) problem, irrespective of \(U_1-U_2\). The switching between $P_L$ and $P_R$ in Fig. \ref{fig:switch} therefore
requires the open, interacting character of the DQD.\\

Moreover, the mechanism underlying the switching is intrinsically many-body physics. The interaction asymmetry ($U_1\neq U_2$) becomes relevant when the DQD is coupled to reservoirs, which allow the number of electrons on the DQD to fluctuate. Although $v_1\simeq v_2$, the two dots have different many-body addition spectra since $U_1\neq U_2$. In particular, the energies required to add a second electron of opposite spin to create onsite double occupation are approximately \(\Delta E_{\mathrm{add}}^{(1)}\simeq v_1+U_1\) for dot 1 and \(\Delta E_{\mathrm{add}}^{(2)}\simeq v_2+U_2\) for dot 2. Thus, for \(v_1\simeq v_2\) but \(U_1\neq U_2\), the dots are equivalent in the $N=1$ sector but inequivalent in the neighboring $N>1$ sectors. A uniform $M_z$ further changes the spin-resolved occupations and the accessibility of the addition and removal channels connecting the dominant \(N=1\) sector to the neighboring \(N=0\) and \(N=2\) sectors. Since the energies of the intermediate \(N=2\) configurations depend differently on \(U_1\) and \(U_2\), these reservoir-mediated processes need not affect the two dots equally. Reservoir coupling therefore
communicates the interaction asymmetry of the higher-charge configurations
to the predominantly single-electron manifold, converting a spatially
uniform Zeeman splitting into a left-right charge response.\\

The resulting switching is illustrated at $v_1=v_2=-1.2$ in
Figs. \ref{fig:switch}(c) and \ref{fig:switch}(d), where $P_L$ and $P_R$ vary strongly with
$M_z$ and cross as the field is swept while $P_{N=1}=P_L+P_R\simeq1$ remains pinned. The magnetic field therefore
reverses the left-right charge polarization while leaving the total DQD charge intact. 
Coulomb blockade is essential to the character of this response. Without sufficiently strong interactions, the Zeeman-dependent exchange of electrons with the reservoirs would predominantly change the total DQD occupation. Large onsite and interdot repulsions suppress additional charge configurations and stabilize the single-electron regime, such that $P_{N=1}=P_L+P_R\simeq1$. The reservoir-induced response then appears primarily as a redistribution of the existing single-electron probability between the dots rather than as a substantial change in the total charge.\\


Importantly, \(P_{N=1}\simeq1\) does not imply that neighboring charge sectors are dynamically irrelevant. Small stationary populations outside the \(N=1\) sector, together with real or virtual processes such as \(N=1\rightarrow N=0\rightarrow N=1\) and \(N=1\rightarrow N=2\rightarrow N=1\), can influence how the dominant single-electron probability is divided between the dots. The present finite-coupling approach does not unambiguously separate the effects of finite stationary occupation of neighboring charge sectors from those of virtual excursions through them. Both arise because dot--reservoir tunnelling allows the charge of the DQD subsystem to fluctuate. These processes probe the interaction-asymmetric addition spectrum and influence how the dominant \(N=1\) probability is distributed between the dots. Importantly, the charge-polarization reversal persists even at $t_{12}=0$,
establishing that the switching is an intrinsic consequence of the
interaction-asymmetric many-body mechanism and does not rely on
tunneling-induced interdot coherence. These tests support the robustness of the predicted redistribution within the density-sector EOM approximation.\\

\textit{Potential applications ---} The practical interest of this mechanism lies in providing a control axis for the charge polarization that is distinct from conventional electrostatic detuning. Ordinarily, the electron is moved between the dots by changing \(v_1-v_2\), which directly modifies the DQD energy landscape and can also affect the alignment with the reservoirs, the tunnel barriers, and neighboring devices through gate cross-talk. Here, the left-right polarization \(D=P_L-P_R\) can instead be varied through $M_z$ while the bare dot levels and their detuning remain fixed. The Zeeman field can therefore reconfigure the spatial distribution of the electron without requiring a large excursion from the chosen electrostatic operating point and without substantially changing the total DQD occupation. Within the dominant \(N=1\) sector, the left- and right-localized configurations provide two distinguishable single-electron charge states ($P_L$ and $P_R$) which may serve as the basis for a charge-qubit encoding. A sign reversal of $D$ changes which dot carries the larger fraction of the single-electron weight and may therefore be useful for state initialization and spin-to-charge conversion.  The resulting redistribution can be detected by a local charge sensor or microwave resonator and can modify the coupling to reservoirs, neighboring quantum dots, or other circuit elements that couple differently to the two sites. A uniform magnetic field could further provide a global control parameter, while local gate voltages select which DQDs are tuned into a regime of strong Zeeman response. On the other hand, Zeeman control may also offer a conditional noise advantage. Because the redistribution does not require a large detuning pulse, it may reduce control-induced charge noise, capacitive cross-talk, and unintended changes in the dot--reservoir alignment. Of particular interest would be an operating point at which \(D\) remains sensitive to \(M_z\) but is first-order insensitive to fluctuations of the electrical control parameters. Experimentally, the interdot Coulomb interaction can be strongly enhanced
and tuned through device geometry and electrostatic gating \cite{PhysRevLett.101.186804,Chan2002,Banszerus2021}.
While the conventional electrostatic capacitance model imposes
$U_{12}<\sqrt{U_1U_2}$ \cite{RevModPhys.75.1}, it is not a general constraint on the interaction parameters of an
effective DQD Hamiltonian \cite{nishino2016exact,sobrinoprb24,coll2021density}. Importantly, switching persists
at reduced $U_{12}$ satisfying $U_{12}<\sqrt{U_1U_2}$ (see Supplemental Material \cite{SM}), establishing
its accessibility in conventional capacitively coupled DQDs.\\

\textit{Conclusion---} 
We have shown that a uniform Zeeman splitting can be converted into spatial charge
polarization through the asymmetric response of an open DQD: gate-level
asymmetry yields pronounced charge reconfiguration, while interaction
asymmetry enables genuine left-right switching. The systematic consistency
tests including  normalization, positivity, symmetry, and limiting-case
checks \cite{SM}, together with the
persistence of switching at vanishing interdot tunneling, establish the
robustness of our symmetry-preserving EOM truncation. This Zeeman-driven
control of left-right charge states provides a new route toward manipulating
DQD charge qubits for quantum computation.

\begin{acknowledgments}
We thank Nahual Sobrino for useful discussions. This work was supported by the Research
Council of Norway through Grant No. 353894 and its Centres
of Excellence funding scheme Grant No. 262633 “QuSpin.” Support from
Sigma2 - the National Infrastructure for High Performance
Computing and Data Storage in Norway, project NN9577K, is acknowledged.
\end{acknowledgments}

\bibliography{bib}

\begin{widetext}
\appendix
\section{DETAILS OF THE GREEN'S-FUNCTION EQUATION-OF-MOTION APPROACH}\label{sec:I}

\subsection{Hamiltonian of the system}

The Hamiltonian of the double quantum dot (DQD) system is given by
\begin{equation}
\hat H = \hat H_{\mathrm{leads}} + \hat H_{\mathrm{2dot}} + \hat H_{\mathrm{tun}},
\end{equation}
where $\hat H_{\mathrm{leads}}$, $\hat H_{\mathrm{2dot}}$ and $\hat H_{\mathrm{tun}}$ represent the Hamiltonian of the free electrons in the leads under thermal equilibrium, the Hamiltonian of the double quantum dots (DQDs), and the tunneling Hamiltonian between the dots and leads, respectively. We have 
\begin{equation}
\hat H_{\mathrm{leads}}
=
\sum_{ik\sigma}
\epsilon_{ik\sigma}
\hat{c}_{ik\sigma}^{\dagger}
\hat{c}_{ik\sigma},
\end{equation}
\begin{equation}
\hat H_{\mathrm{2dot}}
=
\sum_{i\sigma}
v_{i\sigma}
\hat{d}_{i\sigma}^{\dagger}
\hat{d}_{i\sigma}
+
\sum_i
U_i \hat{n}_{i{\uparrow}}\hat{n}_{i{\downarrow}}
+
U_{12}\hat{n}_1\hat{n}_2
+
t_{12}\sum_\sigma
\left(
\hat{d}_{1\sigma}^{\dagger}\hat{d}_{2\sigma}
+\mathrm{H.c.}
\right),
\end{equation}
\begin{equation}
\hat H_{\mathrm{tun}}
=
\sum_{ik\sigma}
\left(
V_{ik\sigma}\hat{c}_{ik\sigma}^{\dagger}\hat{d}_{i\sigma}
+\mathrm{H.c.}
\right),
\end{equation}
in which $\hat{d}_{i\sigma}$ is the annihilation operator of an electron with spin $\sigma$ at QD $i$, and $\hat{c}_{ik\sigma}$ is the annihilation operator of an electron with spin $\sigma$ at momentum state $k$ in the reservoir connected to QD $i$. The occupation operator is
$\hat{n}_i=\sum_\sigma\hat{n}_{i\sigma}$ with
$\hat{n}_{i\sigma}=\hat{d}_{i\sigma}^{\dagger}\hat{d}_{i\sigma}$.
$\epsilon_{ik\sigma}$ describes the single-particle eigen-energies of the reservoirs.
$v_{i\sigma}=v_i+\sigma M_z$ is the QD onsite energy with
$M_z=g^{*}\mu_B B_z$ being the Zeeman energy induced by the magnetic field
$\mathbf{B}=(0,0,B_z)$, in which $v_i$ is called the dot level or gate energy.
$U_i$ and $U_{12}$ represent intra- and interdot Coulomb repulsions, respectively.
$t_{12}$ is the tunneling amplitude between the two QDs.
$V_{ik\sigma}$ denotes the tunneling between the QD and its corresponding lead. In the DQD system, $\bar{i}$ is used to represent the other QD site.

\subsection{Single particle Green's functions}
Define the retarded single-particle Green's function (GF) in the time domain and its Fourier transform as
\begin{align}
    G_{i\sigma}^r(t)&=-i\Theta(t)\langle\{\hat{d}_{i\sigma}(t),\hat{d}_{i\sigma}^\dag(0)\}\rangle,\\
    G_{i\sigma}^r(\omega)&\equiv\langle\langle \hat{d}_{i\sigma}:\hat{d}_{i\sigma}^\dag\rangle\rangle_{\omega}=\int_{-\infty}^{+\infty}G_{i\sigma}^r(t)e^{i\omega t}dt.
\end{align}
It is found that $G_{i\sigma}^r(\omega)$ can be related to the occupation via 
\begin{equation}
    \langle\hat{n}_{i\sigma}\rangle=-2\int  f(\omega)\text{Im}{[G_{i\sigma}^r(\omega)]},
    \label{eq:dev}
\end{equation}
in which $f(\omega)=[1+e^{\beta\hbar\omega}]^{-1}$ with $\beta\equiv1/T$ is the Fermi distribution and $\int\equiv\int\frac{d\omega}{2\pi}$. The derivation details of Eq. (\ref{eq:dev}) can be found in the last section of the Supplementary Material.\\

On the other hand, we also introduce the general retarded GF for a general operator $\hat{B}(t)$ as 
\begin{align}
     G_{B}^r(t)&=-i\Theta(t)\langle\{\hat{B}(t),\hat{d}_{i\sigma}^\dag(0)\}\rangle,\\
    G_{B}^r(\omega)&\equiv\langle\langle \hat{B}:\hat{d}_{i\sigma}^\dag\rangle\rangle_{\omega}=\int_{-\infty}^{+\infty}G_{B}^r(t)e^{i\omega t}dt.
    \label{eq:def_B}
\end{align}
To derive the equation of motion (EOM) of $G_{B}^r(\omega)$, we first calculate the dynamics in the time domain as
\begin{equation}
   \frac{d}{dt}G_{B}^r(t)=-i\frac{d\Theta(t)}{dt}\langle\{\hat{B}(t),\hat{d}_{i\sigma}^\dag(0)\}\rangle-i\Theta(t)\langle\{\frac{d}{dt}\hat{B}(t),\hat{d}_{i\sigma}^\dag(0)\}\rangle.
\end{equation}
Insert the following useful relations 
\begin{equation}
    \frac{d\Theta(t)}{dt}=\delta(t),\qquad \frac{d}{dt}\hat{B}(t)=\frac{1}{i\hbar}[\hat{B}(t),\hat{H}(t)],
\end{equation}
we arrive at
\begin{equation}
    \frac{d}{dt}G_{B}^r(t)=-i\delta(t)\langle\{\hat{B}(t),\hat{d}_{i\sigma}^\dag(0)\}\rangle-i\Theta(t)\langle\{\frac{1}{i\hbar}[\hat{B}(t),\hat{H}(t)],\hat{d}_{i\sigma}^\dag(0)\}\rangle.
\end{equation}
We then perform Fourier transform on the LHS:
\begin{equation}
    \int_{-\infty}^{+\infty}e^{i\omega t}\frac{dG_{B}^r(t)}{dt} dt=e^{i(\omega+i\eta) t}G_{B}^r(t)|_{-\infty}^{+\infty}-\int_{-\infty}^{+\infty}i\omega e^{i\omega t}G_{B}^r(t)dt=-i\omega G_B^r(\omega),
\end{equation}
where $\eta\rightarrow0_+$ is introduced for convergence. On the RHS, by using the integration property for $\delta(t)$ and the general definition of $\langle\langle...\rangle\rangle_\omega$ in Eq. (\ref{eq:def_B}) for the Fourier transform, we get 
\begin{equation}
   -i\omega G_B^r(\omega)=-i\langle\{\hat{B}(0),\hat{d}_{i\sigma}^\dag(0)\}\rangle+\frac{1}{i\hbar}\langle\langle[\hat{B},\hat{H}]:\hat{d}_{i\sigma}^\dag\rangle\rangle_\omega. 
\end{equation}
Rearrange the above equation, we rewrite the EOM of $G_B^r(\omega)$ as
\begin{equation}
   \hbar\omega_{+} G_B^r(\omega)=\hbar\langle\{\hat{B}(0),\hat{d}_{i\sigma}^\dag(0)\}\rangle+\langle\langle[\hat{B}(t),\hat{H}(t)]:\hat{d}_{i\sigma}^\dag(0)\rangle\rangle_\omega,
   \label{eq:EOM_B_new}
\end{equation}
in which  $\omega_{+}\equiv\omega+i\eta$ with $\eta\rightarrow0_+$.\\

To get the EOM of the single particle GF $G_{i\sigma}^r(\omega)$, we simply apply $\hat{B}(t)=\hat{d}_{i\sigma}(t)$ in Eq. (\ref{eq:EOM_B_new}) and get
\begin{align}
    \hbar\omega_{+} G_{i\sigma}^r(\omega)&=\hbar\langle\{\hat{d}_{i\sigma}(0),\hat{d}_{i\sigma}^\dag(0)\}\rangle+\langle\langle[\hat{d}_{i\sigma}(t),\hat{H}(t)]:\hat{d}_{i\sigma}^\dag(0)\rangle\rangle_\omega\notag\\
    &=\hbar+\langle\langle[\hat{d}_{i\sigma}(t),\hat{H}(t)]:\hat{d}_{i\sigma}^\dag(0)\rangle\rangle_\omega\notag\\
    &=\hbar+\langle\langle[\hat{d}_{i\sigma},\hat{H}]:\hat{d}_{i\sigma}^\dag\rangle\rangle_\omega
   \label{eq:EOM_d1_new}
\end{align}
in which $\{\hat{d}_{i\sigma}(0),\hat{d}_{i\sigma}^\dag(0)\}=1$ is utilized and the time-dependence is omitted for simplify in the last step. We then focus on the commutation relation $[\hat{d}_{i\sigma},\hat{H}]$ in the second term on the RHS of Eq. (\ref{eq:EOM_d1_new}). We obtain
\begin{align}
    [\hat{d}_{i\sigma},\hat{H}]&=(v_{i\sigma}+U_i\hat{n}_{i\bar{\sigma}}+U_{12}\hat{n}_{\bar{i}})\hat{d}_{i\sigma}+t_{12}\hat{d}_{\bar{i}\sigma}+\sum_k V_{ik\sigma}^*\hat{c}_{ik\sigma}.
    \label{eq:dH_new}
\end{align}
Insert Eq. (\ref{eq:dH_new}) into Eq. (\ref{eq:EOM_d1_new}), we arrive at
\begin{align}
    (\hbar\omega_+ - v_{i\sigma})\hat{G}_{i\sigma}^r(\omega)&=\hbar+U_i\langle\langle\hat{n}_{i\bar{\sigma}}\hat{d}_{i\sigma}:\hat{d}_{i\sigma}^\dag\rangle\rangle_\omega+U_{12}\langle\langle\hat{n}_{\bar{i}}\hat{d}_{i\sigma}:\hat{d}_{i\sigma}^\dag\rangle\rangle_\omega\notag\\&+t_{12}\hat{G}_{\bar{i}{\sigma}}^r(\omega)+\sum_k V_{i k\sigma}^*\langle\langle\hat{c}_{i k\sigma}:\hat{d}_{i\sigma}^\dag\rangle\rangle_\omega,
    \label{eq:all_GF_new}
\end{align}
in which two single GFs $\hat{G}_{i\sigma}^r(\omega)=\langle\langle \hat{d}_{i\sigma}:\hat{d}_{i\sigma}^\dag\rangle\rangle_{\omega}$ and $\hat{G}_{\bar{i}\sigma}^r(\omega)=\langle\langle \hat{d}_{\bar{i}\sigma}:\hat{d}_{i\sigma}^\dag\rangle\rangle_{\omega}$ appear simultaneously. To get the mixed dot-lead GF in Eq. (\ref{eq:all_GF_new}), i.e., $\langle\langle \hat{c}:\hat{d}^\dag\rangle\rangle_\omega$,
we simply apply $\hat{B}(t)=\hat{c}_{i k\sigma}(t)$ in Eq. (\ref{eq:EOM_B_new}) and get
\begin{align}
    \hbar\omega_{+}\langle\langle\hat{c}_{ik\sigma}:\hat{d}_{i\sigma}^\dag\rangle\rangle_\omega &=\hbar\langle\{\hat{c}_{i k\sigma}(0),\hat{d}_{i\sigma}^\dag(0)\}\rangle+\langle\langle[\hat{c}_{i k\sigma}(t),\hat{H}(t)]:\hat{d}_{i\sigma}^\dag(0)\rangle\rangle_\omega\notag\\&=\langle\langle[\hat{c}_{i k\sigma}(t),\hat{H}(t)]:\hat{d}_{i\sigma}^\dag(0)\rangle\rangle_\omega,
    \label{eq:cHd_new}
\end{align}
where $\{\hat{c}(0),\hat{d}^\dag(0)\}=0$ is utilized. By using the commutation relations
\begin{align}
    [\hat{c}_{i k\sigma},\hat{c}^{\dag}_{j k{'}\sigma^{'}}\hat{c}_{j k{'}\sigma^{'}}]&=\delta_{ij}\delta_{kk^{'}}\delta_{\sigma\sigma^{'}}\hat{c}_{i\alpha k\sigma},\notag\\ 
    [\hat{c}_{i k\sigma},\hat{c}^{\dag}_{j k{'}\sigma^{'}}\hat{d}_{j\sigma^{'}}]&=\delta_{ij}\delta_{kk^{'}}\delta_{\sigma\sigma^{'}}\hat{d}_{i\sigma},\notag\\
    [\hat{c}_{i k\sigma},\hat{d}^\dag_{j\sigma^{'}}\hat{c}_{j k{'}\sigma^{'}}]&=0,
\end{align}
we have 
\begin{align}
    [\hat{c}_{i k\sigma},\hat{H}]&=\epsilon_{i k\sigma}\hat{c}_{i k\sigma} + V_{i k\sigma}\hat{d}_{i\sigma},
    \label{eq:cH_new}
\end{align}
Insert Eq. (\ref{eq:cH_new}) into Eq. (\ref{eq:cHd_new}), we get
\begin{align}
 (\hbar\omega_{+}-\epsilon_{i k\sigma})\langle\langle\hat{c}_{i k\sigma}:\hat{d}_{i\sigma}^\dag\rangle\rangle_\omega&=V_{i k\sigma}G_{i\sigma}^r(\omega).   
\end{align}
Consequently, the last term in Eq. (\ref{eq:all_GF_new}) becomes
\begin{align}
    \sum_k V_{i k\sigma}^*\langle\langle\hat{c}_{ik\sigma}:\hat{d}_{i\sigma}^\dag\rangle\rangle_\omega&=\sum_k V_{i k\sigma}^*\frac{V_{i k\sigma} G_{i\sigma}^r(\omega)}{\hbar\omega_{+}-\epsilon_{i k\sigma}}=\Delta_{i\sigma}(\omega)G_{i\sigma}^r(\omega),
    \label{eq:DeltaG_new}
\end{align}
in which the embedding self-energy or hybridization function $\Delta_{i\sigma}(\omega)$ is defined by
\begin{equation}
\Delta_{i\sigma}(\omega)\equiv\sum_k\frac{|V_{ik\sigma}|^2}{\hbar\omega_{+}-\epsilon_{i k\sigma}},
\label{eq:Delta_new}
\end{equation}
Insert Eq. (\ref{eq:DeltaG_new}) into Eq. (\ref{eq:all_GF_new}), the single particle GF can be written as 
\begin{align}
    [\hbar\omega_+ - v_{i\sigma} - \Delta_{i\sigma}(\omega)]\hat{G}_{i\sigma}^r(\omega)&=\hbar+U_i\langle\langle\hat{n}_{i\bar{\sigma}}\hat{d}_{i\sigma}:\hat{d}_{i\sigma}^\dag\rangle\rangle_\omega+U_{12}\langle\langle\hat{n}_{\bar{i}}\hat{d}_{i\sigma}:\hat{d}_{i\sigma}^\dag\rangle\rangle_\omega\notag\\&+t_{12}\hat{G}_{\bar{i}{\sigma}}^r(\omega),
    \label{eq:GF2_new_new}
\end{align}

Then we need to focus on the remaining high order GFs (i.e., $\langle\langle\hat{n}\hat{d}:\hat{d}^\dag\rangle\rangle_\omega$) on the RHS of Eq. (\ref{eq:GF2_new_new}).

\subsection{Higher-order Green's functions}
As shown in Eq. (\ref{eq:GF2_new_new}), we have two types of single particle GFs $\hat{G}_{i\sigma}^r(\omega)$ and $\hat{G}_{\bar{i}\sigma}^r(\omega)$. Consequently, here we also involve the similar two types of higher-order GFs: $\langle\langle\hat{D}\hat{d}_{i\sigma}:\hat{d}_{i\sigma}^\dag\rangle\rangle_\omega$ and $\langle\langle\hat{D}\hat{d}_{\bar{i}\sigma}:\hat{d}_{i\sigma}^\dag\rangle\rangle_\omega$ with $\hat{D}$ being the product of $N$ density operators.

\subsubsection{$\langle\langle\hat{D}\hat{d}_{i\sigma}:\hat{d}_{i\sigma}^\dag\rangle\rangle_\omega$}\label{sec:Ddd}
We define the general $N+1$ particle dot GF as $G_{Dd_{i\sigma}}^r\equiv\langle\langle\hat{D}\hat{d}_{i\sigma}:\hat{d}_{i\sigma}^\dag\rangle\rangle_\omega$ with $\hat{D}=\hat{n}_{i_1\sigma_1}\hat{n}_{i_2\sigma_2}...\hat{n}_{i_N\sigma_N}$ being the product of $N$ density operators with $i_k\sigma_k\neq i\sigma$ for all $k\in\{1,2,...,N\}$, thus satisfying $[\hat{D},\hat{d}_{i\sigma}]=[\hat{D},\hat{d}_{i\sigma}^\dag]=0$. For $N=0$, we have the single particle with $\hat{D}=\hat{1}$ being the unit operator, which gives the single particle GF $\hat{G}_{i\sigma}^r(\omega)$.\\

To get the EOM of $\langle\langle\hat{D}\hat{d}_{i\sigma}:\hat{d}_{i\sigma}^\dag\rangle\rangle_\omega$, we apply $\hat{B}=\hat{D}\hat{d}_{i\sigma}$ in Eq. (\ref{eq:EOM_B_new}) and get
\begin{align}
    \hbar\omega_{+} \langle\langle\hat{D}\hat{d}_{i\sigma}:\hat{d}_{i\sigma}^\dag\rangle\rangle_\omega&=\hbar\langle\{\hat{D}\hat{d}_{i\sigma},\hat{d}_{i\sigma}^\dag\}\rangle+\langle\langle[\hat{D}\hat{d}_{i\sigma},\hat{H}]:\hat{d}_{i\sigma}^\dag\rangle\rangle_\omega,\notag\\&=\hbar\langle\hat{D}\rangle+\langle\langle[\hat{D}\hat{d}_{i\sigma},\hat{H}]:\hat{d}_{i\sigma}^\dag\rangle\rangle_\omega,
    \label{eq:D1_new}
\end{align}
in which $[\hat{D},\hat{d}_{i\sigma}^\dag]=0$ and $\{\hat{d}_{i\sigma},\hat{d}_{i\sigma}^\dag\}=1$ are utilized to get $\{\hat{D}\hat{d}_{i\sigma},\hat{d}_{i\sigma}^\dag\}=\hat{D}$. We then focus on the commutation $[\hat{D}\hat{d}_{i\sigma},\hat{H}]$ in the 2nd term on the RHS of Eq. (\ref{eq:D1_new}):
\begin{align}
    [\hat{D}\hat{d}_{i\sigma},\hat{H}]&=\hat{D}[\hat{d}_{i\sigma},\hat{H}]+[\hat{D},\hat{H}]\hat{d}_{i\sigma}\approx\hat{D}[\hat{d}_{i\sigma},\hat{H}],
    \label{eq:D2_new}
\end{align}
where approximation $[\hat{D},\hat{H}]\approx0$ (e.g., $[\hat{n}_{i\sigma},\hat{H}]\approx0$) is made for negligible $V_{i\alpha k}$ and much smaller $t_{12}$ compared with the Coulomb repulsions. This assumes the local density operator and the Hamiltonian commute, which becomes exact in the
uncontacted situation $\hat{D}=\hat{1}$. 
The approximation does not correspond to discarding all terms of a
given order in $V_{ik\sigma}$ or $t_{12}$. Rather, the two terms describe
different classes of correlations. The first term governs the
propagation of the electron appearing explicitly in the GF and therefore
retains both dot--reservoir and interdot tunneling. The second describes
the tunneling-induced dynamics of the density product $\hat D$ itself;
it generates additional dot--reservoir and interdot-coherence operators
outside the density-correlator hierarchy retained here. We neglect these
additional dynamical fluctuations and treat the occupations entering
$\hat D$ as quasistatic during the single-particle propagation. For the
dot--reservoir contribution, this is the weak-coupling EOM closure of
Ref. \cite{sobrinoprb24}; here the same closure is extended to finite interdot tunneling.
It becomes exact in the atomic limit $V_{ik\sigma},t_{12}\rightarrow0$
and is applied below in the interaction-dominated regime.
$[\hat{d}_{i\sigma},\hat{H}]$ is then computed for a two-particle Hamiltonian in Eq. (\ref{eq:dH_new}). Insert Eq. (\ref{eq:dH_new}) into Eq. (\ref{eq:D2_new}), we obtain
\begin{align}
    [\hat{D}\hat{d}_{i\sigma},\hat{H}]&=(v_{i\sigma}\hat{D}+U_i\hat{D}\hat{n}_{i\bar{\sigma}}+U_{12}\hat{D}\hat{n}_{\bar{i}})\hat{d}_{i\sigma}\notag\\
    &+t_{12}\hat{D}\hat{d}_{\bar{i}\sigma}+\sum_k V_{ik\sigma}^*\hat{D}\hat{c}_{ik\sigma}.
    \label{eq:D3_new}
\end{align}
Insert Eq. (\ref{eq:D3_new}) into Eq. (\ref{eq:D1_new}), we get
\begin{align}
    (\hbar\omega_+ - v_{i\sigma})\langle\langle\hat{D}\hat{d}_{i\sigma}:\hat{d}_{i\sigma}^\dag\rangle\rangle_\omega&=\hbar\langle\hat{D}\rangle+U_i\langle\langle\hat{D}\hat{n}_{i\bar{\sigma}}\hat{d}_{i\sigma}:\hat{d}_{i\sigma}^\dag\rangle\rangle_\omega+U_{12}\langle\langle\hat{D}\hat{n}_{\bar{i}}\hat{d}_{i\sigma}:\hat{d}_{i\sigma}^\dag\rangle\rangle_\omega\notag\\&+t_{12}\langle\langle\hat{D}\hat{d}_{\bar{i}\sigma}:\hat{d}_{i\sigma}^\dag\rangle\rangle_\omega+\sum_k V_{i k\sigma}^*\langle\langle\hat{D}\hat{c}_{i k\sigma}:\hat{d}_{i\sigma}^\dag\rangle\rangle_\omega.
    \label{eq:D4_new}
\end{align}\\
To cover the higher order mixed dot-lead GF $\langle\langle\hat{D}\hat{c}_{i k\sigma} :\hat{d}_{i\sigma}^\dag\rangle\rangle_\omega$ appearing in the last line of Eq. (\ref{eq:D4_new}), we apply $\hat{B}=\hat{D}\hat{c}_{i k\sigma}$ in Eq. (\ref{eq:EOM_B_new}) and get
\begin{align}
    \hbar\omega_{+} \langle\langle\hat{D}\hat{c}_{i k\sigma}:\hat{d}_{i\sigma}^\dag\rangle\rangle_\omega&=\hbar\langle\{\hat{D}\hat{c}_{i k\sigma},\hat{d}_{i\sigma}^\dag\}\rangle+\langle\langle[\hat{D}\hat{c}_{ik\sigma},\hat{H}]:\hat{d}_{i\sigma}^\dag\rangle\rangle_\omega,\notag\\&=\langle\langle[\hat{D}\hat{c}_{i k\sigma},\hat{H}]:\hat{d}_{i\sigma}^\dag\rangle\rangle_\omega.
    \label{eq:D5_new_1}
\end{align}
The commutation $[\hat{D}\hat{c}_{i k\sigma},\hat{H}]$ in Eq. (\ref{eq:D5_new_1}) is calculated as
\begin{align}
    [\hat{D}\hat{c}_{ik\sigma},\hat{H}]&=\hat{D}[\hat{c}_{i k\sigma},\hat{H}]+[\hat{D},\hat{H}]\hat{c}_{i k\sigma}\approx\hat{D}[\hat{c}_{i k\sigma},\hat{H}],
    \label{eq:D6_new}
\end{align}
in which the approximation $[\hat{D},\hat{H}]\approx0$ is again applied. Insert Eq. (\ref{eq:cH_new}) into Eq. (\ref{eq:D6_new}), we have
\begin{align}
    [\hat{D}\hat{c}_{i k\sigma},\hat{H}]&=\epsilon_{i k\sigma}\hat{D}\hat{c}_{i k\sigma} + V_{i k\sigma}\hat{D}\hat{d}_{i\sigma},
    \label{eq:D7_new}
\end{align}
Insert Eq. (\ref{eq:D7_new}) into Eq. (\ref{eq:D5_new_1}), we get
\begin{align}
 (\hbar\omega_{+}-\epsilon_{i k\sigma})\langle\langle\hat{D}\hat{c}_{i k\sigma}:\hat{d}_{i\sigma}^\dag\rangle\rangle_\omega&=V_{i k\sigma}\langle\langle\hat{D}\hat{d}_{i\sigma}:\hat{d}_{i\sigma}^\dag\rangle\rangle_\omega. 
 \label{eq:D8_new}
b\end{align}
Insert Eq. (\ref{eq:D8_new}) into Eq. (\ref{eq:D4_new}), we obtain

\begin{align}
    [\hbar\omega_+ - v_{i\sigma}-\Delta_{i\sigma}(\omega)]\langle\langle\hat{D}\hat{d}_{i\sigma}:\hat{d}_{i\sigma}^\dag\rangle\rangle_\omega&=\hbar\langle\hat{D}\rangle+U_i\langle\langle\hat{D}\hat{n}_{i\bar{\sigma}}\hat{d}_{i\sigma}:\hat{d}_{i\sigma}^\dag\rangle\rangle_\omega+U_{12}\langle\langle\hat{D}\hat{n}_{\bar{i}}\hat{d}_{i\sigma}:\hat{d}_{i\sigma}^\dag\rangle\rangle_\omega\notag\\&+t_{12}\langle\langle\hat{D}\hat{d}_{\bar{i}\sigma}:\hat{d}_{i\sigma}^\dag\rangle\rangle_\omega.
    \label{eq:GF_D_new}
\end{align}
in which $\Delta_{i\sigma}(\omega)$ is the hybridization function given by Eq. (\ref{eq:Delta_new}).  Note the higher order GF in Eq. (\ref{eq:GF_D_new}) reduces to the single particle GF $\hat{G}_{i\sigma}^r(\omega)$ in Eq. (\ref{eq:GF2_new_new}) with $\hat{D}=\hat{1}$, in which the approximation $[\hat{D},\hat{H}]=0$ becomes exact.\\

Apply $\hat{D}=\hat{1}$ in Eq. (\ref{eq:GF_D_new}), we get
\begin{align}
    \langle\langle \hat{d}_{i\sigma}:\hat{d}_{i\sigma}^\dag\rangle\rangle_\omega&=\frac{\hbar}{\hbar\omega_+ - v_{i\sigma}-\Delta_{i\sigma}(\omega)}+\frac{U_i\langle\langle\hat{n}_{i\bar{\sigma}}\hat{d}_{i\sigma}:\hat{d}_{i\sigma}^\dag\rangle\rangle_\omega}{\hbar\omega_+ - v_{i\sigma}-\Delta_{i\sigma}(\omega)}+\frac{U_{12}\langle\langle\hat{n}_{\bar{i}}\hat{d}_{i\sigma}:\hat{d}_{i\sigma}^\dag\rangle\rangle_\omega}{\hbar\omega_+ - v_{i\sigma}-\Delta_{i\sigma}(\omega)}\notag\\
    &+\frac{t_{12}\langle\langle\hat{d}_{\bar{i}{\sigma}}:\hat{d}_{i\sigma}^\dag\rangle\rangle_\omega}{\hbar\omega_+ - v_{i\sigma}-\Delta_{i\sigma}(\omega)}.
    \label{eq:Dn0_new}
\end{align}
Apply $\hat{D}=\hat{n}_{i\bar{\sigma}}$ in Eq. (\ref{eq:GF_D_new}), we get
\begin{align}
    [\hbar\omega_+ - v_{i\sigma}-\Delta_{i\sigma}(\omega)]\langle\langle\hat{n}_{i\bar{\sigma}}\hat{d}_{i\sigma}:\hat{d}_{i\sigma}^\dag\rangle\rangle_\omega&=\hbar\langle\hat{n}_{i\bar{\sigma}}\rangle+U_i\langle\langle\hat{n}_{i\bar{\sigma}}\hat{n}_{i\bar{\sigma}}\hat{d}_{i\sigma}:\hat{d}_{i\sigma}^\dag\rangle\rangle_\omega+U_{12}\langle\langle\hat{n}_{i\bar{\sigma}}\hat{n}_{\bar{i}}\hat{d}_{i\sigma}:\hat{d}_{i\sigma}^\dag\rangle\rangle_\omega\notag\\&+t_{12}\langle\langle\hat{n}_{i\bar{\sigma}}\hat{d}_{\bar{i}\sigma}:\hat{d}_{i\sigma}^\dag\rangle\rangle_\omega.
    \label{eq:Dn1_new}
\end{align}
We then use the useful relation  $\hat{n}_{i{\sigma}}\hat{n}_{i{\sigma}}=\hat{n}_{i{\sigma}}$ for fermions, which is proved as follows: 
\begin{equation}
    \hat{n}^2=\hat{n}\hat{n}=(\hat{d}^\dag\hat{d})(\hat{d}^\dag\hat{d})=\hat{d}^\dag(\hat{d}\hat{d}^\dag)\hat{d}=\hat{d}^\dag(1-\hat{d}^\dag\hat{d})\hat{d}=\hat{d}^\dag\hat{d}-\hat{d}^\dag\hat{d}^\dag\hat{d}\hat{d}=\hat{d}^\dag\hat{d}=\hat{n},
\end{equation}
in which $\hat{d}^\dag\hat{d}^\dag=0$ is utilized. This expresses the binary  occupancy for fermions according to the Pauli exclusion principle. Consequently, Eq. (\ref{eq:Dn1_new}) becomes
\begin{align}
    \langle\langle \hat{n}_{i\bar{\sigma}}\hat{d}_{i\sigma}:\hat{d}_{i\sigma}^\dag\rangle\rangle_\omega&=\frac{\hbar\langle\hat{n}_{i\bar{\sigma}}\rangle}{\hbar\omega_+ - v_{i\sigma}-U_i-\Delta_{i\sigma}(\omega)}+\frac{U_{12}\langle\langle\hat{n}_{i\bar{\sigma}}\hat{n}_{\bar{i}}\hat{d}_{i\sigma}:\hat{d}_{i\sigma}^\dag\rangle\rangle_\omega}{\hbar\omega_+ - v_{i\sigma}-U_i-\Delta_{i\sigma}(\omega)}\notag\\
    &+\frac{t_{12}\langle\langle\hat{n}_{i\bar{\sigma}}\hat{d}_{\bar{i}{\sigma}}:\hat{d}_{i\sigma}^\dag\rangle\rangle_\omega}{\hbar\omega_+ - v_{i\sigma}-U_i-\Delta_{i\sigma}(\omega)}.
    \label{eq:Dn1_new_new}
\end{align}

Apply $\hat{D}=\hat{n}_{\bar{i}\sigma}$ in Eq. (\ref{eq:GF_D_new}), we get
\begin{align}
    \langle\langle \hat{n}_{\bar{i}\sigma}\hat{d}_{i\sigma}:\hat{d}_{i\sigma}^\dag\rangle\rangle_\omega&=\frac{\hbar\langle\hat{n}_{\bar{i}\sigma}\rangle}{\hbar\omega_+ - v_{i\sigma}-U_{12}-\Delta_{i\sigma}(\omega)}+\frac{U_i\langle\langle\hat{n}_{\bar{i}\sigma}\hat{n}_{i\bar{\sigma}}\hat{d}_{i\sigma}:\hat{d}_{i\sigma}^\dag\rangle\rangle_\omega}{\hbar\omega_+ - v_{i\sigma}-U_{12}-\Delta_{i\sigma}(\omega)}\notag\\&+\frac{U_{12}\langle\langle\hat{n}_{\bar{i}\sigma}\hat{n}_{\bar{i}\bar{\sigma}}\hat{d}_{i\sigma}:\hat{d}_{i\sigma}^\dag\rangle\rangle_\omega}{\hbar\omega_+ - v_{i\sigma}-U_{12}-\Delta_{i\sigma}(\omega)},
    \label{eq:Dn2_new}
\end{align}
in which $\hat{n}_{i\sigma}\hat{d}_{i\sigma}=0$ is utilized.
Apply $\hat{D}=\hat{n}_{\bar{i}\bar{\sigma}}$ in Eq. (\ref{eq:GF_D_new}), we get
\begin{align}
    \langle\langle \hat{n}_{\bar{i}\bar{\sigma}}\hat{d}_{i\sigma}:\hat{d}_{i\sigma}^\dag\rangle\rangle_\omega&=\frac{\hbar\langle\hat{n}_{\bar{i}\bar{\sigma}}\rangle}{\hbar\omega_+ - v_{i\sigma}-U_{12}-\Delta_{i\sigma}(\omega)}+\frac{U_i\langle\langle\hat{n}_{\bar{i}\bar{\sigma}}\hat{n}_{i\bar{\sigma}}\hat{d}_{i\sigma}:\hat{d}_{i\sigma}^\dag\rangle\rangle_\omega}{\hbar\omega_+ - v_{i\sigma}-U_{12}-\Delta_{i\sigma}(\omega)}\notag\\&+\frac{U_{12}\langle\langle\hat{n}_{\bar{i}\bar{\sigma}}\hat{n}_{\bar{i}\sigma}\hat{d}_{i\sigma}:\hat{d}_{i\sigma}^\dag\rangle\rangle_\omega}{\hbar\omega_+ - v_{i\sigma}-U_{12}-\Delta_{i\sigma}(\omega)}+\frac{t_{12}\langle\langle\hat{n}_{\bar{i}\bar{\sigma}}\hat{d}_{\bar{i}{\sigma}}:\hat{d}_{i\sigma}^\dag\rangle\rangle_\omega}{\hbar\omega_+ - v_{i\sigma}-U_{12}-\Delta_{i\sigma}(\omega)}.
    \label{eq:Dn3_new}
\end{align}
Similarly, the three three-body GFs are obtained by applying $\hat{D}=\hat{n}_{i\bar{\sigma}}\hat{n}_{\bar{i}\sigma}$, $\hat{n}_{i\bar{\sigma}}\hat{n}_{\bar{i}\bar{\sigma}}$ and $\hat{n}_{\bar{i}{\sigma}}\hat{n}_{\bar{i}\bar{\sigma}}$ in Eq. (\ref{eq:GF_D_new}). For $\hat{D}=\hat{n}_{i\bar{\sigma}}\hat{n}_{\bar{i}\sigma}$ in Eq. (\ref{eq:GF_D_new}), we have
\begin{align}
    \langle\langle \hat{n}_{i\bar{\sigma}}\hat{n}_{\bar{i}\sigma}\hat{d}_{i\sigma}:\hat{d}_{i\sigma}^\dag\rangle\rangle_\omega&=\frac{\hbar\langle\hat{n}_{i\bar{\sigma}}\hat{n}_{\bar{i}\sigma}\rangle}{\hbar\omega_+ - v_{i\sigma}-U_i-U_{12}-\Delta_{i\sigma}(\omega)}+\frac{U_{12}\langle\langle\hat{n}_{i\bar{\sigma}}\hat{n}_{\bar{i}\sigma}\hat{n}_{\bar{i}\bar{\sigma}}\hat{d}_{i\sigma}:\hat{d}_{i\sigma}^\dag\rangle\rangle_\omega}{\hbar\omega_+ - v_{i\sigma}-U_i-U_{12}-\Delta_{i\sigma}(\omega)}.
    \label{eq:Dnn1_new}
\end{align}
For $\hat{D}=\hat{n}_{i\bar{\sigma}}\hat{n}_{\bar{i}\bar{\sigma}}$ in Eq. (\ref{eq:GF_D_new}), we have
\begin{align}
    \langle\langle \hat{n}_{i\bar{\sigma}}\hat{n}_{\bar{i}\bar{\sigma}}\hat{d}_{i\sigma}:\hat{d}_{i\sigma}^\dag\rangle\rangle_\omega&=\frac{\hbar\langle\hat{n}_{i\bar{\sigma}}\hat{n}_{\bar{i}\bar{\sigma}}\rangle}{\hbar\omega_+ - v_{i\sigma}-U_i-U_{12}-\Delta_{i\sigma}(\omega)}+\frac{U_{12}\langle\langle\hat{n}_{i\bar{\sigma}}\hat{n}_{\bar{i}\bar{\sigma}}\hat{n}_{\bar{i}\sigma}\hat{d}_{i\sigma}:\hat{d}_{i\sigma}^\dag\rangle\rangle_\omega}{\hbar\omega_+ - v_{i\sigma}-U_i-U_{12}-\Delta_{i\sigma}(\omega)}\notag\\
    &+\frac{t_{12}\langle\langle\hat{n}_{i\bar{\sigma}}\hat{n}_{\bar{i}\bar{\sigma}}\hat{d}_{\bar{i}{\sigma}}:\hat{d}_{i\sigma}^\dag\rangle\rangle_\omega}{\hbar\omega_+ - v_{i\sigma}-U_i-U_{12}-\Delta_{i\sigma}(\omega)}.
    \label{eq:Dnn2_new}
\end{align}
For $\hat{D}=\hat{n}_{\bar{i}{\sigma}}\hat{n}_{\bar{i}\bar{\sigma}}$ in Eq. (\ref{eq:GF_D_new}), we have
\begin{align}
    \langle\langle \hat{n}_{\bar{i}{\sigma}}\hat{n}_{\bar{i}\bar{\sigma}}\hat{d}_{i\sigma}:\hat{d}_{i\sigma}^\dag\rangle\rangle_\omega&=\frac{\hbar\langle\hat{n}_{\bar{i}{\sigma}}\hat{n}_{\bar{i}\bar{\sigma}}\rangle}{\hbar\omega_+ - v_{i\sigma}-2U_{12}-\Delta_{i\sigma}(\omega)}+\frac{U_i\langle\langle\hat{n}_{\bar{i}{\sigma}}\hat{n}_{\bar{i}\bar{\sigma}}\hat{n}_{i\bar{\sigma}}\hat{d}_{i\sigma}:\hat{d}_{i\sigma}^\dag\rangle\rangle_\omega}{\hbar\omega_+ - v_{i\sigma}-2U_{12}-\Delta_{i\sigma}(\omega)}.
    \label{eq:Dnn3_new}
\end{align}
Finally, the four-body GF is determined by applying $\hat{D}=\hat{n}_{i\bar{\sigma}}\hat{n}_{\bar{i}\sigma}\hat{n}_{\bar{i}\bar{\sigma}}$ as
\begin{align}
    \langle\langle \hat{n}_{i\bar{\sigma}}\hat{n}_{\bar{i}\sigma}\hat{n}_{\bar{i}\bar{\sigma}}\hat{d}_{i\sigma}:\hat{d}_{i\sigma}^\dag\rangle\rangle_\omega&=\frac{\hbar\langle\hat{n}_{i\bar{\sigma}}\hat{n}_{\bar{i}\sigma}\hat{n}_{\bar{i}\bar{\sigma}}\rangle}{\hbar\omega_+ - v_{i\sigma}-U_i-2U_{12}-\Delta_{i\sigma}(\omega)},
    \label{eq:Dnnn1_new}
\end{align}
which is the highest-order GF in the EOM hierarchy. \\

\subsubsection{$\langle\langle\hat{D}\hat{d}_{\bar{i}\sigma}:\hat{d}_{i\sigma}^\dag\rangle\rangle_\omega$}\label{sec:sub}
Following the same procedure for $G_{Dd_{{i}\sigma}}^r$, we on the other hand define the general $N+1$ particle dot GF as $G_{Dd_{\bar{i}\sigma}}^r\equiv\langle\langle\hat{D}\hat{d}_{\bar{i}\sigma}:\hat{d}_{i\sigma}^\dag\rangle\rangle_\omega$ with $\hat{D}=\hat{n}_{i_1\sigma_1}\hat{n}_{i_2\sigma_2}...\hat{n}_{i_N\sigma_N}$ being the product of $N$ density operators with $i_k\sigma_k\neq \bar{i}\sigma$ for all $k\in\{1,2,...,N\}$, thus satisfying $[\hat{D},\hat{d}_{\bar{i}\sigma}]=[\hat{D},\hat{d}_{\bar{i}\sigma}^\dag]=0$. For $N=0$, we have the single particle with $\hat{D}=\hat{1}$ being the unit operator, which gives the single particle GF $\hat{G}_{\bar{i}\sigma}^r(\omega)$.\\

To get the EOM of $\langle\langle\hat{D}\hat{d}_{\bar{i}\sigma}:\hat{d}_{i\sigma}^\dag\rangle\rangle_\omega$, we apply $\hat{B}=\hat{D}\hat{d}_{\bar{i}\sigma}$ in Eq. (\ref{eq:EOM_B_new}) and get
\begin{align}
    \hbar\omega_{+} \langle\langle\hat{D}\hat{d}_{\bar{i}\sigma}:\hat{d}_{i\sigma}^\dag\rangle\rangle_\omega&=\hbar\langle\{\hat{D}\hat{d}_{\bar{i}\sigma},\hat{d}_{i\sigma}^\dag\}\rangle+\langle\langle[\hat{D}\hat{d}_{\bar{i}\sigma},\hat{H}]:\hat{d}_{i\sigma}^\dag\rangle\rangle_\omega,\notag\\&=\langle\langle[\hat{D}\hat{d}_{\bar{i}\sigma},\hat{H}]:\hat{d}_{i\sigma}^\dag\rangle\rangle_\omega,
    \label{eq:D1_new_a1}
\end{align}
in which $[\hat{D},\hat{d}_{\bar{i}\sigma}^\dag]=0$ and $\{\hat{d}_{\bar{i}\sigma},\hat{d}_{i\sigma}^\dag\}=0$ are utilized to get $\{\hat{D}\hat{d}_{\bar{i}\sigma},\hat{d}_{i\sigma}^\dag\}=0$ when $\hat{D}$ does not contain $\hat{n}_{i\sigma}$. As for $\hat D$ contains $\hat n_{i\sigma}$, the anticommutator instead
generates an interdot-coherence $\langle\hat d^\dagger_{i\sigma}\hat d_{\bar i\sigma}\rangle$ or density-assisted interdot-coherence [$\langle\hat{F}\hat d^\dagger_{i\sigma}\hat d_{\bar i\sigma}\rangle$ with $\hat D=\hat n_{i\sigma}\hat F$]. These off-diagonal contributions are consistently omitted
within the same density-correlator closure as the approximation
$[\hat D,\hat H]\simeq0$, as discussed below.  \\

The approximation $[\hat D,\hat H]\simeq0$ should be understood as a
closure of the EOM within the occupation-diagonal density-correlator
sector. For finite $t_{12}$, the omitted commutator
$[\hat D,\hat H_{12}]$ with $\hat{H}_{12}=t_{12}\sum_\sigma
\left(
\hat{d}_{1\sigma}^{\dagger}\hat{d}_{2\sigma}
+\mathrm{H.c.}
\right)$ generates interdot-coherence ($\hat d^\dagger_{i\sigma}\hat d_{\bar i\sigma}$) and
density-assisted-coherence operators ($\hat{F}\hat d^\dagger_{i\sigma}\hat d_{\bar i\sigma}$ with $\hat D=\hat n_{i\sigma}\hat F$). The same coherence sector appears
in the inhomogeneous term of the mixed GF
$\langle\!\langle \hat D\hat d_{\bar i\sigma}:
\hat d^\dagger_{i\sigma}\rangle\!\rangle_\omega$.
Specifically, when $\hat D=\hat n_{i\sigma}\hat F$, with $\hat F$
denoting the product of the remaining density operators in $\hat D$,
\begin{equation}
\{
\hat D\hat d_{\bar i\sigma},
\hat d^\dagger_{i\sigma}
\}
=
-\hat F\hat d^\dagger_{i\sigma}\hat d_{\bar i\sigma}.
\end{equation}
This density-assisted interdot coherence belongs to the same
tunneling-induced off-diagonal sector generated by
$[\hat D,\hat H_{12}]$. Consistent with the approximation
$[\hat D,\hat H]\simeq0$, we therefore project out these coherence
contributions and retain only the density-correlator hierarchy. \\

Importantly, neglecting these interdot coherence variables does not amount to neglecting interdot tunneling. The direct effect of $t_{12}$ on the propagation and consequently on the density correlators is retained explicitly through $\hat D[\hat d_{i\sigma},\hat H]$. What is neglected is the additional dynamical evolution of the off-diagonal coherence sector generated by
$[\hat D,\hat H_{12}]$. This density-sector closure is appropriate for the interaction-dominated
regime considered here, where $t_{12}$ and the dot--reservoir coupling
are smaller than the Coulomb energy scales and the quantities of
interest are the correlated occupation probabilities and their
redistribution. The approximation therefore retains the interaction-
dependent many-body population dynamics and the direct influence of
finite interdot tunneling{and interdot Coulomb interaction}, while neglecting the additional off-diagonal
coherence dynamics that would require an enlarged EOM hierarchy.\\

We then focus on the commutation $[\hat{D}\hat{d}_{\bar{i}\sigma},\hat{H}]$ on the RHS:
\begin{align}
    [\hat{D}\hat{d}_{\bar{i}\sigma},\hat{H}]&=\hat{D}[\hat{d}_{\bar{i}\sigma},\hat{H}]+[\hat{D},\hat{H}]\hat{d}_{\bar{i}\sigma}\approx\hat{D}[\hat{d}_{\bar{i}\sigma},\hat{H}],
    \label{eq:D2_new_a1}
\end{align}
in which the approximation $[\hat{D},\hat{H}]\approx0$ (i.e., $[\hat{n}_{i\sigma},\hat{H}]\approx0$) is made. By replacing $i\rightarrow\bar{i}$ in Eq. (\ref{eq:dH_new}), we obtain
\begin{align}
    [\hat{d}_{\bar{i}\sigma},\hat{H}]&=(v_{\bar{i}\sigma}+U_{\bar{i}}\hat{n}_{\bar{i}\bar{\sigma}}+U_{12}\hat{n}_{{i}})\hat{d}_{\bar{i}\sigma}+t_{12}\hat{d}_{{i}\sigma}+\sum_k V_{\bar{i}k\sigma}^*\hat{c}_{\bar{i}k\sigma}.
    \label{eq:dH_new_a1}
\end{align}
Insert Eq. (\ref{eq:dH_new_a1}) into Eq. (\ref{eq:D2_new_a1}), we obtain
\begin{align}
[\hat{D}\hat{d}_{\bar{i}\sigma},\hat{H}]&=(v_{\bar{i}\sigma}\hat{D}+U_{\bar{i}}\hat{D}\hat{n}_{\bar{i}\bar{\sigma}}+U_{12}\hat{D}\hat{n}_{{i}})\hat{d}_{\bar{i}\sigma}+t_{12}\hat{D}\hat{d}_{{i}\sigma}+\sum_k V_{\bar{i}k\sigma}^*\hat{D}\hat{c}_{\bar{i}k\sigma}.
    \label{eq:D3_new_a1}
\end{align}
Insert Eq. (\ref{eq:D3_new_a1}) into Eq. (\ref{eq:D1_new_a1}), we get
\begin{align}
    (\hbar\omega_+ - v_{\bar{i}{\sigma}})\langle\langle\hat{D}\hat{d}_{\bar{i}{\sigma}}:\hat{d}_{i\sigma}^\dag\rangle\rangle_\omega&=U_{\bar{i}}\langle\langle\hat{D}\hat{n}_{\bar{i}\bar{\sigma}}\hat{d}_{\bar{i}{\sigma}}:\hat{d}_{i\sigma}^\dag\rangle\rangle_\omega+U_{12}\langle\langle\hat{D}\hat{n}_{{i}}\hat{d}_{\bar{i}{\sigma}}:\hat{d}_{i\sigma}^\dag\rangle\rangle_\omega\notag\\&+t_{12}\langle\langle\hat{D}\hat{d}_{{i}{\sigma}}:\hat{d}_{i\sigma}^\dag\rangle\rangle_\omega+\sum_k V_{\bar{i} k{\sigma}}^*\langle\langle\hat{D}\hat{c}_{\bar{i} k{\sigma}}:\hat{d}_{i\sigma}^\dag\rangle\rangle_\omega.
    \label{eq:D4_new_a1}
\end{align}
The higher order mixed dot-lead GFs $\langle\langle\hat{D}\hat{c}_{\bar{i} k{\sigma}} :\hat{d}_{i\sigma}^\dag\rangle\rangle_\omega$ is given by
\begin{align}
 (\hbar\omega_{+}-\epsilon_{\bar{i} k{\sigma}})\langle\langle\hat{D}\hat{c}_{\bar{i} k{\sigma}}:\hat{d}_{i\sigma}^\dag\rangle\rangle_\omega&=V_{\bar{i} k{\sigma}}\langle\langle\hat{D}\hat{d}_{\bar{i}{\sigma}}:\hat{d}_{i\sigma}^\dag\rangle\rangle_\omega,
\label{eq:D8_new_a1}
\end{align}
Insert Eq. (\ref{eq:D8_new_a1}) into Eq. (\ref{eq:D4_new_a1}), we obtain
\begin{align}
    [\hbar\omega_+ - v_{\bar{i}{\sigma}}-\Delta_{\bar{i}{\sigma}}(\omega)]\langle\langle\hat{D}\hat{d}_{\bar{i}\sigma}:\hat{d}_{i\sigma}^\dag\rangle\rangle_\omega&=U_{\bar{i}}\langle\langle\hat{D}\hat{n}_{\bar{i}\bar{\sigma}}\hat{d}_{\bar{i}{\sigma}}:\hat{d}_{i\sigma}^\dag\rangle\rangle_\omega+U_{12}\langle\langle\hat{D}\hat{n}_{{i}}\hat{d}_{\bar{i}{\sigma}}:\hat{d}_{i\sigma}^\dag\rangle\rangle_\omega\notag\\&+t_{12}\langle\langle\hat{D}\hat{d}_{{i}{\sigma}}:\hat{d}_{i\sigma}^\dag\rangle\rangle_\omega.
   \label{eq:GF_D_new_a1}
\end{align}

in which $\Delta_{i\sigma}(\omega)$ is given by Eq. (\ref{eq:Delta_new}). Note here we have $\Delta_{\bar{i}\sigma}(\omega)$ instead.\\

Apply $\hat{D}=\hat{1}$ in Eq. (\ref{eq:GF_D_new_a1}), we get
\begin{align}
    \langle\langle \hat{d}_{\bar{i}{\sigma}}:\hat{d}_{i\sigma}^\dag\rangle\rangle_\omega&=\frac{U_{\bar{i}}\langle\langle\hat{n}_{\bar{i}\bar{\sigma}}\hat{d}_{\bar{i}{\sigma}}:\hat{d}_{i\sigma}^\dag\rangle\rangle_\omega}{\hbar\omega_+ - v_{\bar{i}{\sigma}}-\Delta_{\bar{i}{\sigma}}(\omega)}+\frac{U_{12}\langle\langle\hat{n}_{{i}}\hat{d}_{\bar{i}{\sigma}}:\hat{d}_{i\sigma}^\dag\rangle\rangle_\omega}{\hbar\omega_+ - v_{\bar{i}{\sigma}}-\Delta_{\bar{i}{\sigma}}(\omega)}\notag\\
    &+\frac{t_{12}\langle\langle\hat{d}_{{i}{\sigma}}:\hat{d}_{i\sigma}^\dag\rangle\rangle_\omega}{\hbar\omega_+ - v_{\bar{i}{\sigma}}-\Delta_{\bar{i}{\sigma}}(\omega)}.
    \label{eq:Dn0_new_a1}
\end{align}
Apply $\hat{D}=\hat{n}_{i\sigma}$ in Eq. (\ref{eq:GF_D_new_a1}), we get
\begin{align}
    \langle\langle \hat{n}_{i\sigma}\hat{d}_{\bar{i}{\sigma}}:\hat{d}_{i\sigma}^\dag\rangle\rangle_\omega&=\frac{U_{\bar{i}}\langle\langle\hat{n}_{i\sigma}\hat{n}_{\bar{i}\bar{\sigma}}\hat{d}_{\bar{i}{\sigma}}:\hat{d}_{i\sigma}^\dag\rangle\rangle_\omega}{\hbar\omega_+ - v_{\bar{i}{\sigma}}-U_{12}-\Delta_{\bar{i}{\sigma}}(\omega)}+\frac{U_{12}\langle\langle\hat{n}_{i\sigma}\hat{n}_{{i}\bar{\sigma}}\hat{d}_{\bar{i}{\sigma}}:\hat{d}_{i\sigma}^\dag\rangle\rangle_\omega}{\hbar\omega_+ - v_{\bar{i}{\sigma}}-U_{12}-\Delta_{\bar{i}{\sigma}}(\omega)}.
    \label{eq:Dn1_new_a1}
\end{align}

Apply $\hat{D}=\hat{n}_{{i}\bar{\sigma}}$ in Eq. (\ref{eq:GF_D_new_a1}), we get
\begin{align}
    \langle\langle \hat{n}_{{i}\bar{\sigma}}\hat{d}_{\bar{i}{\sigma}}:\hat{d}_{i\sigma}^\dag\rangle\rangle_\omega&=\frac{U_{\bar{i}}\langle\langle\hat{n}_{{i}\bar{\sigma}}\hat{n}_{\bar{i}\bar{\sigma}}\hat{d}_{\bar{i}{\sigma}}:\hat{d}_{i\sigma}^\dag\rangle\rangle_\omega}{\hbar\omega_+ - v_{\bar{i}{\sigma}}-U_{12}-\Delta_{\bar{i}{\sigma}}(\omega)}+\frac{U_{12}\langle\langle\hat{n}_{{i}\bar{\sigma}}\hat{n}_{{i}\sigma}\hat{d}_{\bar{i}{\sigma}}:\hat{d}_{i\sigma}^\dag\rangle\rangle_\omega}{\hbar\omega_+ - v_{\bar{i}{\sigma}}-U_{12}-\Delta_{\bar{i}{\sigma}}(\omega)}\notag\\
    &+\frac{t_{12}\langle\langle\hat{n}_{{i}\bar{\sigma}}\hat{d}_{{i}{\sigma}}:\hat{d}_{i\sigma}^\dag\rangle\rangle_\omega}{\hbar\omega_+ - v_{\bar{i}{\sigma}}-U_{12}-\Delta_{\bar{i}{\sigma}}(\omega)}.
    \label{eq:Dn2_new_a1}
\end{align}

Apply $\hat{D}=\hat{n}_{\bar{i}\bar{\sigma}}$ in Eq. (\ref{eq:GF_D_new_a1}), we get
\begin{align}
    \langle\langle \hat{n}_{\bar{i}\bar{\sigma}}\hat{d}_{\bar{i}{\sigma}}:\hat{d}_{i\sigma}^\dag\rangle\rangle_\omega&=\frac{U_{12}\langle\langle\hat{n}_{\bar{i}\bar{\sigma}}\hat{n}_{{i}}\hat{d}_{\bar{i}{\sigma}}:\hat{d}_{i\sigma}^\dag\rangle\rangle_\omega}{\hbar\omega_+ - v_{\bar{i}{\sigma}}-U_{\bar{i}}-\Delta_{\bar{i}{\sigma}}(\omega)}+\frac{t_{12}\langle\langle\hat{n}_{\bar{i}\bar{\sigma}}\hat{d}_{{i}{\sigma}}:\hat{d}_{i\sigma}^\dag\rangle\rangle_\omega}{\hbar\omega_+ - v_{\bar{i}{\sigma}}-U_{\bar{i}}-\Delta_{\bar{i}{\sigma}}(\omega)}.
    \label{eq:Dn3_new_a1}
\end{align}
Apply $\hat{D}=\hat{n}_{i\sigma}\hat{n}_{{i}\bar{\sigma}}$ in Eq. (\ref{eq:GF_D_new_a1}), 
\begin{align}
    \langle\langle \hat{n}_{i\sigma}\hat{n}_{{i}\bar{\sigma}}\hat{d}_{\bar{i}{\sigma}}:\hat{d}_{i\sigma}^\dag\rangle\rangle_\omega&=\frac{U_{\bar{i}}\langle\langle\hat{n}_{i\sigma}\hat{n}_{{i}\bar{\sigma}}\hat{n}_{\bar{i}\bar{\sigma}}\hat{d}_{\bar{i}{\sigma}}:\hat{d}_{i\sigma}^\dag\rangle\rangle_\omega}{\hbar\omega_+ - v_{\bar{i}{\sigma}}-2U_{12}-\Delta_{\bar{i}{\sigma}}(\omega)}.
    \label{eq:Dnn1_new_a1}
\end{align}

Apply $\hat{D}=\hat{n}_{i\sigma}\hat{n}_{\bar{i}\bar{\sigma}}$ in Eq. (\ref{eq:GF_D_new_a1}),
\begin{align}
    \langle\langle \hat{n}_{i\sigma}\hat{n}_{\bar{i}\bar{\sigma}}\hat{d}_{\bar{i}{\sigma}}:\hat{d}_{i\sigma}^\dag\rangle\rangle_\omega&=\frac{U_{12}\langle\langle\hat{n}_{i\sigma}\hat{n}_{\bar{i}\bar{\sigma}}\hat{n}_{{i}\bar{\sigma}}\hat{d}_{\bar{i}{\sigma}}:\hat{d}_{i\sigma}^\dag\rangle\rangle_\omega}{\hbar\omega_+ - v_{\bar{i}{\sigma}}-U_{\bar{i}}-U_{12}-\Delta_{\bar{i}{\sigma}}(\omega)}.
    \label{eq:Dnn2_new_a1}
\end{align}
Apply $\hat{D}=\hat{n}_{{i}\bar{\sigma}}\hat{n}_{\bar{i}\bar{\sigma}}$ in Eq. (\ref{eq:GF_D_new_a1}), 
\begin{align}
    \langle\langle \hat{n}_{{i}\bar{\sigma}}\hat{n}_{\bar{i}\bar{\sigma}}\hat{d}_{\bar{i}{\sigma}}:\hat{d}_{i\sigma}^\dag\rangle\rangle_\omega&=\frac{U_{12}\langle\langle\hat{n}_{{i}\bar{\sigma}}\hat{n}_{\bar{i}\bar{\sigma}}\hat{n}_{{i}\sigma}\hat{d}_{\bar{i}{\sigma}}:\hat{d}_{i\sigma}^\dag\rangle\rangle_\omega}{\hbar\omega_+ - v_{\bar{i}{\sigma}}-U_{\bar{i}}-U_{12}-\Delta_{\bar{i}{\sigma}}(\omega)}+\frac{t_{12}\langle\langle\hat{n}_{{i}\bar{\sigma}}\hat{n}_{\bar{i}\bar{\sigma}}\hat{d}_{{i}{\sigma}}:\hat{d}_{i\sigma}^\dag\rangle\rangle_\omega}{\hbar\omega_+ - v_{\bar{i}{\sigma}}-U_{\bar{i}}-U_{12}-\Delta_{\bar{i}{\sigma}}(\omega)}.
    \label{eq:Dnn3_new_a1}
\end{align}
Apply $\hat{D}=\hat{n}_{{i}\sigma}\hat{n}_{{i}\bar{\sigma}}\hat{n}_{\bar{i}\bar{\sigma}}$ in Eq. (\ref{eq:GF_D_new_a1}),
\begin{align}
    \langle\langle \hat{n}_{{i}\sigma}\hat{n}_{{i}\bar{\sigma}}\hat{n}_{\bar{i}\bar{\sigma}}\hat{d}_{\bar{i}{\sigma}}:\hat{d}_{i\sigma}^\dag\rangle\rangle_\omega&=0.\label{eq:nnn00_a1}
\end{align}

\subsection{Derive the single-particle Green's function by iteration}

\subsubsection{Express $\langle\langle\hat{D}\hat{d}_{\bar{i}{\sigma}}:\hat{d}_{i\sigma}^\dag\rangle\rangle_\omega$ in terms of $\langle\langle\hat{D}\hat{d}_{i\sigma}:\hat{d}_{i\sigma}^\dag\rangle\rangle_\omega$}

Here we express $\langle\langle\hat{D}\hat{d}_{\bar{i}{\sigma}}:\hat{d}_{i\sigma}^\dag\rangle\rangle_\omega$ in terms of $\langle\langle\hat{D}\hat{d}_{i\sigma}:\hat{d}_{i\sigma}^\dag\rangle\rangle_\omega$ by iteration. In section \ref{sec:sub}, insert the four-body GF into the three-body GFs, we obtain
\begin{align}
   \langle\langle \hat{n}_{{i}\bar{\sigma}}\hat{n}_{\bar{i}\bar{\sigma}}\hat{d}_{\bar{i}{\sigma}}:\hat{d}_{i\sigma}^\dag\rangle\rangle_\omega&=\frac{t_{12}\langle\langle\hat{n}_{{i}\bar{\sigma}}\hat{n}_{\bar{i}\bar{\sigma}}\hat{d}_{{i}{\sigma}}:\hat{d}_{i\sigma}^\dag\rangle\rangle_\omega}{\hbar\omega_+ - v_{\bar{i}{\sigma}}-U_{\bar{i}}-U_{12}-\Delta_{\bar{i}{\sigma}}(\omega)},\label{eq:143}\\
   \langle\langle \hat{n}_{i\sigma}\hat{n}_{\bar{i}\bar{\sigma}}\hat{d}_{\bar{i}{\sigma}}:\hat{d}_{i\sigma}^\dag\rangle\rangle_\omega&=0,\\
   \langle\langle \hat{n}_{i\sigma}\hat{n}_{{i}\bar{\sigma}}\hat{d}_{\bar{i}{\sigma}}:\hat{d}_{i\sigma}^\dag\rangle\rangle_\omega&=0.
\end{align}
Insert the above three-body GFs into the two-body GFs in Sec. \ref{sec:sub}, we get
\begin{align}
    \langle\langle \hat{n}_{\bar{i}\bar{\sigma}}\hat{d}_{\bar{i}{\sigma}}:\hat{d}_{i\sigma}^\dag\rangle\rangle_\omega&=[\frac{t_{12}}{\hbar\omega_+ - v_{\bar{i}{\sigma}}-U_{\bar{i}}-U_{12}-\Delta_{\bar{i}{\sigma}}(\omega)}-\frac{t_{12}}{\hbar\omega_+ - v_{\bar{i}{\sigma}}-U_{\bar{i}}-\Delta_{\bar{i}{\sigma}}(\omega)}]\langle\langle\hat{n}_{{i}\bar{\sigma}}\hat{n}_{\bar{i}\bar{\sigma}}\hat{d}_{{i}{\sigma}}:\hat{d}_{i\sigma}^\dag\rangle\rangle_\omega\notag\\&+\frac{t_{12}\langle\langle\hat{n}_{\bar{i}\bar{\sigma}}\hat{d}_{{i}{\sigma}}:\hat{d}_{i\sigma}^\dag\rangle\rangle_\omega}{\hbar\omega_+ - v_{\bar{i}{\sigma}}-U_{\bar{i}}-\Delta_{\bar{i}{\sigma}}(\omega)},\\
    \langle\langle \hat{n}_{{i}\bar{\sigma}}\hat{d}_{\bar{i}{\sigma}}:\hat{d}_{i\sigma}^\dag\rangle\rangle_\omega&=[\frac{t_{12}}{\hbar\omega_+ - v_{\bar{i}{\sigma}}-U_{\bar{i}}-U_{12}-\Delta_{\bar{i}{\sigma}}(\omega)}-\frac{t_{12}}{\hbar\omega_+ - v_{\bar{i}{\sigma}}-U_{12}-\Delta_{\bar{i}{\sigma}}(\omega)}]\langle\langle\hat{n}_{{i}\bar{\sigma}}\hat{n}_{\bar{i}\bar{\sigma}}\hat{d}_{{i}{\sigma}}:\hat{d}_{i\sigma}^\dag\rangle\rangle_\omega\notag\\&+\frac{t_{12}\langle\langle\hat{n}_{{i}\bar{\sigma}}\hat{d}_{{i}{\sigma}}:\hat{d}_{i\sigma}^\dag\rangle\rangle_\omega}{\hbar\omega_+ - v_{\bar{i}{\sigma}}-U_{12}-\Delta_{\bar{i}{\sigma}}(\omega)},\\
    \langle\langle \hat{n}_{i\sigma}\hat{d}_{\bar{i}{\sigma}}:\hat{d}_{i\sigma}^\dag\rangle\rangle_\omega&=0.
\end{align}
Insert the above two-body GFs into the single GF, we have
\begin{align}
    \langle\langle \hat{d}_{\bar{i}{\sigma}}:\hat{d}_{i\sigma}^\dag\rangle\rangle_\omega&=[\frac{t_{12}}{\hbar\omega_+ - v_{\bar{i}{\sigma}}-U_{\bar{i}}-U_{12}-\Delta_{\bar{i}{\sigma}}(\omega)}-\frac{t_{12}}{\hbar\omega_+ - v_{\bar{i}{\sigma}}-U_{\bar{i}}-\Delta_{\bar{i}{\sigma}}(\omega)}\notag\\&-\frac{t_{12}}{\hbar\omega_+ - v_{\bar{i}{\sigma}}-U_{12}-\Delta_{\bar{i}{\sigma}}(\omega)}+\frac{t_{12}}{\hbar\omega_+ - v_{\bar{i}{\sigma}}-\Delta_{\bar{i}{\sigma}}(\omega)}]\langle\langle\hat{n}_{{i}\bar{\sigma}}\hat{n}_{\bar{i}\bar{\sigma}}\hat{d}_{{i}{\sigma}}:\hat{d}_{i\sigma}^\dag\rangle\rangle_\omega\notag\\
    &+[\frac{t_{12}}{\hbar\omega_+ - v_{\bar{i}{\sigma}}-U_{\bar{i}}-\Delta_{\bar{i}{\sigma}}(\omega)}-\frac{t_{12}}{\hbar\omega_+ - v_{\bar{i}{\sigma}}-\Delta_{\bar{i}{\sigma}}(\omega)}]\langle\langle\hat{n}_{\bar{i}\bar{\sigma}}\hat{d}_{{i}{\sigma}}:\hat{d}_{i\sigma}^\dag\rangle\rangle_\omega\notag\\
    &+[\frac{t_{12}}{\hbar\omega_+ - v_{\bar{i}{\sigma}}-U_{12}-\Delta_{\bar{i}{\sigma}}(\omega)}-\frac{t_{12}}{\hbar\omega_+ - v_{\bar{i}{\sigma}}-\Delta_{\bar{i}{\sigma}}(\omega)}]\langle\langle\hat{n}_{{i}\bar{\sigma}}\hat{d}_{{i}{\sigma}}:\hat{d}_{i\sigma}^\dag\rangle\rangle_\omega\notag\\
    &+\frac{t_{12}\langle\langle\hat{d}_{{i}{\sigma}}:\hat{d}_{i\sigma}^\dag\rangle\rangle_\omega}{\hbar\omega_+ - v_{\bar{i}{\sigma}}-\Delta_{\bar{i}{\sigma}}(\omega)}.
\end{align}
Based on the above, it is seen that $\langle\langle\hat{D}\hat{d}_{\bar{i}{\sigma}}:\hat{d}_{i\sigma}^\dag\rangle\rangle_\omega\rightarrow0$ if $t_{12}\rightarrow0$.

\subsubsection{Express $\langle\langle\hat{D}\hat{d}_{i\sigma}:\hat{d}_{i\sigma}^\dag\rangle\rangle_\omega$ in terms of density correlators}

For notation simplicity, we introduce the single-pole notations
$p_{i\sigma,j}$
\begin{align}
    p_{i\sigma,1}=v_{i\sigma},\qquad p_{i\sigma,2}=v_{i\sigma}+U_i,\qquad p_{i\sigma,3}=v_{i\sigma}+U_i+U_{12},\notag\\p_{i\sigma,4}=v_{i\sigma}+U_i+2U_{12},\qquad p_{i\sigma,5}=v_{i\sigma}+U_{12},\qquad p_{i\sigma,6}=v_{i\sigma}+2U_{12}
\end{align}
with $P_{i\sigma,j}=p_{i\sigma,j}+\Delta_{i\sigma}$.\\

In section \ref{sec:Ddd}, insert the four-body GF into the three-body GFs, we get
\begin{align}
     \langle\langle \hat{n}_{\bar{i}{\sigma}}\hat{n}_{\bar{i}\bar{\sigma}}\hat{d}_{i\sigma}:\hat{d}_{i\sigma}^\dag\rangle\rangle_\omega=\frac{\hbar(\langle\hat{n}_{\bar{i}{\sigma}}\hat{n}_{\bar{i}\bar{\sigma}}\rangle-\langle\hat{n}_{i\bar{\sigma}}\hat{n}_{\bar{i}\sigma}\hat{n}_{\bar{i}\bar{\sigma}}\rangle)}{
    \hbar\omega_{+}-P_{i\sigma,6}}+\frac{\hbar\langle\hat{n}_{i\bar{\sigma}}\hat{n}_{\bar{i}\sigma}\hat{n}_{\bar{i}\bar{\sigma}}\rangle}{\hbar\omega_{+}-P_{i\sigma,4}},
    \label{eq:new_nnd1}
\end{align}
\begin{align}
    \langle\langle\hat{n}_{i\bar{\sigma}}\hat{n}_{\bar{i}\sigma} \hat{d}_{i\sigma}:\hat{d}_{i\sigma}^\dag\rangle\rangle_\omega=\frac{\hbar(\langle\hat{n}_{i\bar{\sigma}}\hat{n}_{\bar{i}\sigma}\rangle-\langle\hat{n}_{i\bar{\sigma}}\hat{n}_{\bar{i}\sigma}\hat{n}_{\bar{i}\bar{\sigma}}\rangle)}{
    \hbar\omega_{+}-P_{i\sigma,3}}+\frac{\hbar\langle\hat{n}_{i\bar{\sigma}}\hat{n}_{\bar{i}\sigma}\hat{n}_{\bar{i}\bar{\sigma}}\rangle}{\hbar\omega_{+}-P_{i\sigma,4}}, \label{eq:new_nnd2}
\end{align}
\begin{align}
    \langle\langle \hat{n}_{i\bar{\sigma}}\hat{n}_{\bar{i}\bar{\sigma}}\hat{d}_{i\sigma}:\hat{d}_{i\sigma}^\dag\rangle\rangle_\omega&=\frac{\hbar(\langle\hat{n}_{i\bar{\sigma}}\hat{n}_{\bar{i}\bar{\sigma}}\rangle-\langle\hat{n}_{i\bar{\sigma}}\hat{n}_{\bar{i}\sigma}\hat{n}_{\bar{i}\bar{\sigma}}\rangle)}{
    \hbar\omega_{+}-P_{i\sigma,3}}+\frac{\hbar\langle\hat{n}_{i\bar{\sigma}}\hat{n}_{\bar{i}\sigma}\hat{n}_{\bar{i}\bar{\sigma}}\rangle}{\hbar\omega_{+}-P_{i\sigma,4}}+\frac{t_{12}\langle\langle\hat{n}_{i\bar{\sigma}}\hat{n}_{\bar{i}\bar{\sigma}}\hat{d}_{\bar{i}{\sigma}}:\hat{d}_{i\sigma}^\dag\rangle\rangle_\omega}{\hbar\omega_+ - P_{i\sigma,3}}.
    \label{eq:new_nnd3}
\end{align}
Insert $\langle\langle\hat{n}_{i\bar{\sigma}}\hat{n}_{\bar{i}\bar{\sigma}}\hat{d}_{\bar{i}{\sigma}}:\hat{d}_{i\sigma}^\dag\rangle\rangle_\omega$ given in Eq. (\ref{eq:143}) into Eq. (\ref{eq:new_nnd3}), we rewrite

\begin{align}
    \langle\langle \hat{n}_{i\bar{\sigma}}\hat{n}_{\bar{i}\bar{\sigma}}\hat{d}_{i\sigma}:\hat{d}_{i\sigma}^\dag\rangle\rangle_\omega&=\frac{\hbar(\hbar\omega_+ -P_{\bar{i}\sigma,3})}{(\hbar\omega_+ -P_{i\sigma,3})(\hbar\omega_+ -P_{\bar{i}\sigma,3})-t_{12}^2}\big[\langle\hat{n}_{i\bar{\sigma}}\hat{n}_{\bar{i}\bar{\sigma}}\rangle-\langle\hat{n}_{i\bar{\sigma}}\hat{n}_{\bar{i}\sigma}\hat{n}_{\bar{i}\bar{\sigma}}\rangle+\frac{\hbar\omega_+ -P_{i\sigma,3}}{\hbar\omega_+-P_{i\sigma,4}}\langle\hat{n}_{i\bar{\sigma}}\hat{n}_{\bar{i}\sigma}\hat{n}_{\bar{i}\bar{\sigma}}\rangle \big]\notag\\&=\frac{\hbar(\langle\hat{n}_{i\bar{\sigma}}\hat{n}_{\bar{i}\bar{\sigma}}\rangle-\langle\hat{n}_{i\bar{\sigma}}\hat{n}_{\bar{i}\sigma}\hat{n}_{\bar{i}\bar{\sigma}}\rangle)}{K_{i\sigma,1}(\omega)} + \frac{\hbar\langle\hat{n}_{i\bar{\sigma}}\hat{n}_{\bar{i}\sigma}\hat{n}_{\bar{i}\bar{\sigma}}\rangle}{K_{i\sigma,2}(\omega)},
    \label{eq:n3}
\end{align}
 Now we have hybridized poles determined by $(\hbar\omega_+ -P_{i\sigma,3})(\hbar\omega_+ -P_{\bar{i}\sigma,3})-t_{12}^2=0$. If $t_{12}=0$, this reduces to the case with only single poles. For notation simplicity $K_1(\omega)$ and $K_2(\omega)$ are introduced to describe the hybridized poles with  
 \begin{align}
     K_{i\sigma,1}(\omega)&=\frac{(\hbar\omega_+ -P_{i\sigma,3})(\hbar\omega_+ -P_{\bar{i}\sigma,3})-t_{12}^2}{\hbar\omega_+ -P_{\bar{i}\sigma,3}},\\
     K_{i\sigma,2}(\omega)&=K_{i\sigma,1}\frac{\hbar\omega_+ -P_{{i}\sigma,4}}{\hbar\omega_+ -P_{{i}\sigma,3}}=\frac{(\hbar\omega_+ -P_{i\sigma,3})(\hbar\omega_+ -P_{\bar{i}\sigma,3})-t_{12}^2}{\hbar\omega_+ -P_{\bar{i}\sigma,3}}\frac{\hbar\omega_+ -P_{{i}\sigma,4}}{\hbar\omega_+ -P_{{i}\sigma,3}}.
 \end{align}

Insert the above three-body GF into the two-body GFs, we get
\begin{align}
    \langle\langle \hat{n}_{\bar{i}\sigma}\hat{d}_{i\sigma}:\hat{d}_{i\sigma}^\dag\rangle\rangle_\omega&=\frac{\hbar(\langle\hat{n}_{\bar{i}\sigma}\rangle-\langle\hat{n}_{{i}\bar{\sigma}}\hat{n}_{\bar{i}\sigma}\rangle-\langle\hat{n}_{\bar{i}\sigma}\hat{n}_{\bar{i}\bar{\sigma}}\rangle+\langle\hat{n}_{i\bar{\sigma}}\hat{n}_{\bar{i}\sigma}\hat{n}_{\bar{i}\bar{\sigma}}\rangle)}{
    \hbar\omega_{+}-P_{i\sigma,5}}+\frac{\hbar(\langle\hat{n}_{{i}\bar{\sigma}}\hat{n}_{\bar{i}\sigma}\rangle-\langle\hat{n}_{i\bar{\sigma}}\hat{n}_{\bar{i}\sigma}\hat{n}_{\bar{i}\bar{\sigma}}\rangle)}{\hbar\omega_{+}-P_{i\sigma,3}}\notag\\&+\frac{\hbar(\langle\hat{n}_{\bar{i}{\sigma}}\hat{n}_{\bar{i}\bar{\sigma}}\rangle-\langle\hat{n}_{i\bar{\sigma}}\hat{n}_{\bar{i}\sigma}\hat{n}_{\bar{i}\bar{\sigma}}\rangle)}{\hbar\omega_{+}-P_{i\sigma,6}}+\frac{\hbar\langle\hat{n}_{i\bar{\sigma}}\hat{n}_{\bar{i}\sigma}\hat{n}_{\bar{i}\bar{\sigma}}\rangle}{\hbar\omega_{+}-P_{i\sigma,4}},
    \label{eq:n4}
\end{align}

\begin{align}
    \langle\langle \hat{n}_{i\bar{\sigma}}\hat{d}_{i\sigma}:\hat{d}_{i\sigma}^\dag\rangle\rangle_\omega&=\frac{\hbar}{K_{i\sigma,3}(\omega)}\bigg[\langle\hat{n}_{i\bar{\sigma}}\rangle+\big(\langle\hat{n}_{i\bar{\sigma}}\hat{n}_{\bar{i}{\sigma}}\rangle-\langle\hat{n}_{i\bar{\sigma}}\hat{n}_{\bar{i}\sigma}\hat{n}_{\bar{i}\bar{\sigma}}\rangle\big)\frac{U_{12}}{\hbar\omega_{+}-P_{i\sigma,3}}\notag\\&+\big(\langle\hat{n}_{i\bar{\sigma}}\hat{n}_{\bar{i}\bar{\sigma}}\rangle-\langle\hat{n}_{i\bar{\sigma}}\hat{n}_{\bar{i}\sigma}\hat{n}_{\bar{i}\bar{\sigma}}\rangle\big)\bigg(\frac{U_{12}}{K_{i\sigma,1}(\omega)} + \frac{t_{12}^2 U_{\bar{i}}}{(\hbar\omega_{+}-P_{\bar{i}\sigma,3})(\hbar\omega_{+}-P_{\bar{i}\sigma,5})K_{i\sigma,1}(\omega)}\bigg)\notag\\&+\langle\hat{n}_{i\bar{\sigma}}\hat{n}_{\bar{i}\sigma}\hat{n}_{\bar{i}\bar{\sigma}}\rangle\bigg(\frac{U_{12}}{\hbar\omega_{+}-P_{i\sigma,4}}+\frac{U_{12}}{K_{i\sigma,2}(\omega)}+\frac{t_{12}^2 U_{\bar{i}}}{(\hbar\omega_{+}-P_{\bar{i}\sigma,3})(\hbar\omega_{+}-P_{\bar{i}\sigma,5})K_{i\sigma,2}(\omega)}\bigg)\bigg],
     \label{eq:n55}
\end{align}
\begin{align}
    \langle\langle \hat{n}_{\bar{i}\bar{\sigma}}\hat{d}_{i\sigma}:\hat{d}_{i\sigma}^\dag\rangle\rangle_\omega&=\frac{\hbar}{K_{i\sigma,4}(\omega)}\bigg[\langle\hat{n}_{\bar{i}\bar{\sigma}}\rangle+\big(\langle\hat{n}_{\bar{i}{\sigma}}\hat{n}_{\bar{i}\bar{\sigma}}\rangle-\langle\hat{n}_{i\bar{\sigma}}\hat{n}_{\bar{i}\sigma}\hat{n}_{\bar{i}\bar{\sigma}}\rangle\big)\frac{U_{12}}{\hbar\omega_{+}-P_{i\sigma,6}}\notag\\&\big(\langle\hat{n}_{{i}\bar{\sigma}}\hat{n}_{\bar{i}\bar{\sigma}}\rangle-\langle\hat{n}_{i\bar{\sigma}}\hat{n}_{\bar{i}\sigma}\hat{n}_{\bar{i}\bar{\sigma}}\rangle\big)\bigg(\frac{U_{i}}{K_{i\sigma,1}(\omega)} + \frac{t_{12}^2U_{12}}{(\hbar\omega_+ -P_{\bar{i}\sigma,2})(\hbar\omega_+ -P_{\bar{i}\sigma,3})K_{i\sigma,1}(\omega)}
    \bigg)\notag\\&+\langle\hat{n}_{i\bar{\sigma}}\hat{n}_{\bar{i}\sigma}\hat{n}_{\bar{i}\bar{\sigma}}\rangle\bigg(\frac{U_{12}}{\hbar\omega_{+}-P_{i\sigma,4}} + \frac{U_{i}}{K_{i\sigma,2}(\omega)} + \frac{t_{12}^2U_{12}}{(\hbar\omega_+ -P_{\bar{i}\sigma,2})(\hbar\omega_+ -P_{\bar{i}\sigma,3})K_{i\sigma,2}(\omega)}
    \bigg)\bigg].
     \label{eq:n66}
\end{align}
in which the additional notations $K_{i\sigma,3}$ and $K_{i\sigma,3}$ are introduced to cover hybridized poles
\begin{align}
     K_{i\sigma,3}(\omega)&=\frac{(\hbar\omega_+ -P_{i\sigma,2})(\hbar\omega_+ -P_{\bar{i}\sigma,5})-t_{12}^2}{\hbar\omega_+ -P_{\bar{i}\sigma,5}},\\
     K_{i\sigma,4}(\omega)&=\frac{(\hbar\omega_+ -P_{i\sigma,5})(\hbar\omega_+ -P_{\bar{i}\sigma,2})-t_{12}^2}{\hbar\omega_+ -P_{\bar{i}\sigma,2}}.
 \end{align}
 
Insert the above two-body GFs into the one-body GF, we get
\begin{align}
    \langle\langle \hat{d}_{i\sigma}:\hat{d}_{i\sigma}^\dag\rangle\rangle_\omega/\hbar&=g_{i\sigma,0}+A_{i\sigma,4}\langle\hat{n}_{{i}\bar{\sigma}}\rangle + A_{i\sigma,5}\langle\hat{n}_{\bar{i}{\sigma}}\rangle + A_{i\sigma,6}\langle\hat{n}_{\bar{i}\bar{\sigma}}\rangle\notag\\&+
    B_{i\sigma,4}\langle\hat{n}_{{i}\bar{\sigma}}\hat{n}_{\bar{i}{\sigma}}\rangle + B_{i\sigma,5}\langle\hat{n}_{\bar{i}{\sigma}}\hat{n}_{\bar{i}\bar{\sigma}}\rangle + B_{i\sigma,6}\langle\hat{n}_{{i}\bar{\sigma}}\hat{n}_{\bar{i}\bar{\sigma}}\rangle+C_{i\sigma,4}\langle\hat{n}_{i\bar{\sigma}}\hat{n}_{\bar{i}\sigma}\hat{n}_{\bar{i}\bar{\sigma}}\rangle,
    \label{eq:ddnew}
\end{align}
in which
\begin{equation}
    K_{i\sigma,5}(\omega)=\frac{1}{g_{i\sigma,0}}=\frac{(\hbar\omega_+ -P_{i\sigma,1})(\hbar\omega_+ -P_{\bar{i}\sigma,1})-t_{12}^2}{\hbar\omega_+ -P_{\bar{i}\sigma,1}}.
\end{equation}
The explicit expressions of the remaining coefficients $A_{i\sigma,4}$, $A_{i\sigma,5}$, $A_{i\sigma,6}$, $B_{i\sigma,4}$,$B_{i\sigma,5}$, $B_{i\sigma,6}$ and $C_{i\sigma,4}$ are given in the next section.

\subsection{Linear system for density correlators}
In the previous section, we have expressed GFs in terms of density correlators. On the other hand, the density correlators can also be obtained from the GFs through 
\begin{equation}
    \langle\hat{D}\hat{n}_{i\sigma}\rangle=-2\int f(\omega)\im[G_{Dd_{i\sigma}}^r(\omega)],
    \label{eq:D_to_n_new}
\end{equation}
in which $\int\equiv\int d\omega/(2\pi)$. The derivation details of Eq. (\ref{eq:D_to_n_new}) is given in the last section of the Supplementary Material. By inserting Eqs. (\ref{eq:new_nnd1}, \ref{eq:new_nnd2}, \ref{eq:n3}, \ref{eq:n4}, \ref{eq:n55}, \ref{eq:n66}, \ref{eq:ddnew}) into Eq. (\ref{eq:D_to_n_new}), we can derive a linear system for the
density correlators which can easily be solved numerically.\\

In the wide-band limit (WBL), the hybridiztaion function $\Delta_{i\sigma}(\omega)$ becomes constant and purely imaginary, i.e., $\Delta_{i\sigma}(\omega)=\Delta_{i}=-i\gamma_i/2$ with $\gamma=\gamma_1=\gamma_2$. Note we have only single poles for GFs in Eqs. (\ref{eq:new_nnd1}, \ref{eq:new_nnd2}, \ref{eq:n4}), in which the integral in Eq. (\ref{eq:D_to_n_new}) can be evaluated analytically as
\begin{equation}
    \phi(p)=\int f(\omega)\frac{\gamma}{(\omega-p)^2+(\gamma/2)^2}=\frac{1}{2}-\frac{1}{\pi}\im[\psi(\frac{1}{2}+\frac{\gamma/2+ip}{2\pi T})],
    \label{eq:phi_new}
\end{equation}
where $\int\equiv\int d\omega/(2\pi)$,  $\psi(z)=\frac{d\log(\Gamma(z))}{dz}$ is the digamma function with general complex argument $z$, and $\Gamma(z)$ is the gamma function. Insert Eqs. (\ref{eq:new_nnd1}, \ref{eq:new_nnd2}, \ref{eq:n4}) into Eq. (\ref{eq:D_to_n_new}) and use Eq. (\ref{eq:phi_new}) for the integral, we get
\begin{align}
     \langle \hat{n}_{\bar{i}{\sigma}}\hat{n}_{\bar{i}\bar{\sigma}}\hat{n}_{i\sigma}\rangle/\hbar&=\phi(p_{i\sigma,6})(\langle\hat{n}_{\bar{i}{\sigma}}\hat{n}_{\bar{i}\bar{\sigma}}\rangle-\langle\hat{n}_{i\bar{\sigma}}\hat{n}_{\bar{i}\sigma}\hat{n}_{\bar{i}\bar{\sigma}}\rangle)+\phi(p_{i\sigma,4})\langle\hat{n}_{i\bar{\sigma}}\hat{n}_{\bar{i}\sigma}\hat{n}_{\bar{i}\bar{\sigma}}\rangle,
    \label{eq:lin_1}
\end{align}
\begin{align}
    \langle\hat{n}_{i\bar{\sigma}}\hat{n}_{\bar{i}\sigma} \hat{n}_{i\sigma}\rangle/\hbar=&\phi(p_{i\sigma,3})(\langle\hat{n}_{i\bar{\sigma}}\hat{n}_{\bar{i}\sigma}\rangle-\langle\hat{n}_{i\bar{\sigma}}\hat{n}_{\bar{i}\sigma}\hat{n}_{\bar{i}\bar{\sigma}}\rangle)+\phi(p_{i\sigma,4})\langle\hat{n}_{i\bar{\sigma}}\hat{n}_{\bar{i}\sigma}\hat{n}_{\bar{i}\bar{\sigma}}\rangle, \label{eq:lin_2}
\end{align}
\begin{align}
    \langle \hat{n}_{\bar{i}\sigma}\hat{n}_{i\sigma}\rangle/\hbar&=\phi(p_{i\sigma,5})(\langle\hat{n}_{\bar{i}\sigma}\rangle-\langle\hat{n}_{{i}\bar{\sigma}}\hat{n}_{\bar{i}\sigma}\rangle-\langle\hat{n}_{\bar{i}\sigma}\hat{n}_{\bar{i}\bar{\sigma}}\rangle+\langle\hat{n}_{i\bar{\sigma}}\hat{n}_{\bar{i}\sigma}\hat{n}_{\bar{i}\bar{\sigma}}\rangle)+\phi(p_{i\sigma,3})(\langle\hat{n}_{{i}\bar{\sigma}}\hat{n}_{\bar{i}\sigma}\rangle-\langle\hat{n}_{i\bar{\sigma}}\hat{n}_{\bar{i}\sigma}\hat{n}_{\bar{i}\bar{\sigma}}\rangle)\notag\\&+\phi(p_{i\sigma,6})(\langle\hat{n}_{\bar{i}{\sigma}}\hat{n}_{\bar{i}\bar{\sigma}}\rangle-\langle\hat{n}_{i\bar{\sigma}}\hat{n}_{\bar{i}\sigma}\hat{n}_{\bar{i}\bar{\sigma}}\rangle)+\phi(p_{i\sigma,4})\langle\hat{n}_{i\bar{\sigma}}\hat{n}_{\bar{i}\sigma}\hat{n}_{\bar{i}\bar{\sigma}}\rangle.
    \label{eq:lin_3}
\end{align}\\

As for GFs given by Eqs. (\ref{eq:n3}, \ref{eq:n55}, \ref{eq:n66}, \ref{eq:ddnew}) with hybridized poles, the integral in Eq. (\ref{eq:D_to_n_new}) can only be evaluated numerically. Before perform the integration, we first simplify the coefficient notations in front of the density correlators ($A$, $B$ and $C$ for one-body, two-body and three-body correlators, respectively). We then rewrite
\begin{align}
    \langle\langle \hat{n}_{i\bar{\sigma}}\hat{n}_{\bar{i}\bar{\sigma}}\hat{d}_{i\sigma}:\hat{d}_{i\sigma}^\dag\rangle\rangle_\omega/\hbar=B_{i\sigma,1}\langle\hat{n}_{i\bar{\sigma}}\hat{n}_{\bar{i}\bar{\sigma}}\rangle+C_{i\sigma,1}\langle\hat{n}_{i\bar{\sigma}}\hat{n}_{\bar{i}\sigma}\hat{n}_{\bar{i}\bar{\sigma}}\rangle,
    \label{eq:n33}
\end{align}

\begin{align}
    \langle\langle \hat{n}_{i\bar{\sigma}}\hat{d}_{i\sigma}:\hat{d}_{i\sigma}^\dag\rangle\rangle_\omega/\hbar=A_{i\sigma,2}\langle\hat{n}_{i\bar{\sigma}}\rangle + B_{i\sigma,2}\langle\hat{n}_{i\bar{\sigma}}\hat{n}_{\bar{i}{\sigma}}\rangle + C_{i\sigma,2}\langle\hat{n}_{i\bar{\sigma}}\hat{n}_{\bar{i}\sigma}\hat{n}_{\bar{i}\bar{\sigma}}\rangle+D_{i\sigma,2}\langle\hat{n}_{i\bar{\sigma}}\hat{n}_{\bar{i}\bar{\sigma}}\rangle,
\end{align}

\begin{align}
    \langle\langle \hat{n}_{\bar{i}\bar{\sigma}}\hat{d}_{i\sigma}:\hat{d}_{i\sigma}^\dag\rangle\rangle_\omega/\hbar=A_{i\sigma,3}\langle\hat{n}_{\bar{i}\bar{\sigma}}\rangle + B_{i\sigma,3}\langle\hat{n}_{\bar{i}{\sigma}}\hat{n}_{\bar{i}\bar{\sigma}}\rangle + C_{i\sigma,3}\langle\hat{n}_{i\bar{\sigma}}\hat{n}_{\bar{i}\sigma}\hat{n}_{\bar{i}\bar{\sigma}}\rangle+D_{i\sigma,3}\langle\hat{n}_{{i}\bar{\sigma}}\hat{n}_{\bar{i}\bar{\sigma}}\rangle,
\end{align}

\begin{align}
    \langle\langle \hat{d}_{i\sigma}:\hat{d}_{i\sigma}^\dag\rangle\rangle_\omega/\hbar&=g_{i\sigma,0}+A_{i\sigma,4}\langle\hat{n}_{{i}\bar{\sigma}}\rangle + A_{i\sigma,5}\langle\hat{n}_{\bar{i}{\sigma}}\rangle + A_{i\sigma,6}\langle\hat{n}_{\bar{i}\bar{\sigma}}\rangle\notag\\&+
    B_{i\sigma,4}\langle\hat{n}_{{i}\bar{\sigma}}\hat{n}_{\bar{i}{\sigma}}\rangle + B_{i\sigma,5}\langle\hat{n}_{\bar{i}{\sigma}}\hat{n}_{\bar{i}\bar{\sigma}}\rangle + B_{i\sigma,6}\langle\hat{n}_{{i}\bar{\sigma}}\hat{n}_{\bar{i}\bar{\sigma}}\rangle+C_{i\sigma,4}\langle\hat{n}_{i\bar{\sigma}}\hat{n}_{\bar{i}\sigma}\hat{n}_{\bar{i}\bar{\sigma}}\rangle,
\end{align}
in which
\begin{align}
    B_{i\sigma,1}&=\frac{1}{K_{i\sigma,1}(\omega)},\quad C_{i\sigma,1}=\frac{1}{K_{i\sigma,2}(\omega)}-\frac{1}{K_{i\sigma,1}(\omega)},\quad A_{i\sigma,2}=\frac{1}{K_{i\sigma,3}(\omega)}, \quad A_{i\sigma,3}=\frac{1}{K_{i\sigma,4}(\omega)},\quad g_{i\sigma,0}=\frac{1}{K_{i\sigma,5}(\omega)},\notag\\
    B_{i\sigma,2}&=\frac{1}{K_{i\sigma,3}(\omega)}\frac{U_{12}}{\hbar\omega_{+}-P_{i\sigma,3}},\quad D_{i\sigma,2}=\frac{1}{K_{i\sigma,3}(\omega)}\bigg( \frac{U_{12}}{K_{i\sigma,1}(\omega)} + \frac{t_{12}^2 U_{\bar{i}}}{(\hbar\omega_{+}-P_{\bar{i}\sigma,3})(\hbar\omega_{+}-P_{\bar{i}\sigma,5})K_{i\sigma,1}(\omega)}
    \bigg),\notag\\
    C_{i\sigma,2}&=\frac{1}{K_{i\sigma,3}(\omega)}\bigg[\bigg(\frac{U_{12}}{\hbar\omega_{+}-P_{i\sigma,4}}+\frac{U_{12}}{K_{i\sigma,2}(\omega)}+\frac{t_{12}^2 U_{\bar{i}}}{(\hbar\omega_{+}-P_{\bar{i}\sigma,3})(\hbar\omega_{+}-P_{\bar{i}\sigma,5})K_{i\sigma,2}(\omega)}\bigg)\notag\\&-\bigg(\frac{U_{12}}{\hbar\omega_{+}-P_{i\sigma,3}} + \frac{U_{12}}{K_{i\sigma,1}(\omega)} + \frac{t_{12}^2 U_{\bar{i}}}{(\hbar\omega_{+}-P_{\bar{i}\sigma,3})(\hbar\omega_{+}-P_{\bar{i}\sigma,5})K_{i\sigma,1}(\omega)}
    \bigg)\bigg],\notag\\ B_{i\sigma,3}&=\frac{1}{K_{i\sigma,4}(\omega)}\frac{U_{12}}{\hbar\omega_{+}-P_{i\sigma,6}} ,\quad D_{i\sigma,3}=\frac{1}{K_{i\sigma,4}(\omega)}\bigg(\frac{U_{i}}{K_{i\sigma,1}(\omega)} + \frac{t_{12}^2U_{12}}{(\hbar\omega_+ -P_{\bar{i}\sigma,2})(\hbar\omega_+ -P_{\bar{i}\sigma,3})K_{i\sigma,1}(\omega)}
    \bigg),\notag\\
    C_{i\sigma,3}&=\frac{1}{K_{i\sigma,4}(\omega)}\bigg[\bigg(\frac{U_{12}}{\hbar\omega_{+}-P_{i\sigma,4}} + \frac{U_{i}}{K_{i\sigma,2}(\omega)} + \frac{t_{12}^2U_{12}}{(\hbar\omega_+ -P_{\bar{i}\sigma,2})(\hbar\omega_+ -P_{\bar{i}\sigma,3})K_{i\sigma,2}(\omega)}
    \bigg)\notag\\&-\bigg(\frac{U_{12}}{\hbar\omega_{+}-P_{i\sigma,6}} + \frac{U_{i}}{K_{i\sigma,1}(\omega)} + \frac{t_{12}^2U_{12}}{(\hbar\omega_+ -P_{\bar{i}\sigma,2})(\hbar\omega_+ -P_{\bar{i}\sigma,3})K_{i\sigma,1}(\omega)}
    \bigg)\bigg],\notag\\
A_{i\sigma,4}&=\frac{A_{i\sigma,2}}{K_{i\sigma,5}}[U_i+\frac{U_{12}t_{12}^2}{(\hbar\omega_{+}-P_{\bar{i}\sigma,1})(\hbar\omega_{+}-P_{\bar{i}\sigma,5})}], A_{i\sigma,5}=\frac{U_{12}}{K_{i\sigma,5}(\hbar\omega_{+}-P_{{i}\sigma,5})}, A_{i\sigma,6}=\frac{A_{i\sigma,3}}{K_{i\sigma,5}}[U_{12}+\frac{U_{\bar{i}}t_{12}^2}{(\hbar\omega_{+}-P_{\bar{i}\sigma,2})(\hbar\omega_{+}-P_{\bar{i}\sigma,1})}],\notag\\
B_{i\sigma,4}&=\frac{1}{K_{i\sigma,5}}\bigg[B_{i\sigma,2}\big(U_i+\frac{U_{12}t_{12}^2}{(\hbar\omega_{+}-P_{\bar{i}\sigma,1})(\hbar\omega_{+}-P_{\bar{i}\sigma,5})}\big)+U_{12}\big(\frac{1}{\hbar\omega_{+}-P_{{i}\sigma,3}}-\frac{1}{\hbar\omega_{+}-P_{{i}\sigma,5}}\big)\bigg],\notag\\B_{i\sigma,5}&=\frac{1}{K_{i\sigma,5}}\bigg[B_{i\sigma,3}\big(U_{12}+\frac{U_{\bar{i}}t_{12}^2}{(\hbar\omega_{+}-P_{\bar{i}\sigma,1})(\hbar\omega_{+}-P_{\bar{i}\sigma,2})}\big)+U_{12}\big(\frac{1}{\hbar\omega_{+}-P_{{i}\sigma,6}}-\frac{1}{\hbar\omega_{+}-P_{{i}\sigma,5}}\big)\bigg],\notag\\B_{i\sigma,6}&=\frac{t_{12}^2 B_{i\sigma,1}}{K_{i\sigma,5}}(\frac{1}{\hbar\omega_{+}-P_{\bar{i}\sigma,3}}-\frac{1}{\hbar\omega_{+}-P_{\bar{i}\sigma,2}}-\frac{1}{\hbar\omega_{+}-P_{\bar{i}\sigma,5}}+\frac{1}{\hbar\omega_{+}-P_{\bar{i}\sigma,1}})\notag\\&+\frac{D_{i\sigma,2}}{K_{i\sigma,5}}[U_i+\frac{t_{12}^2 U_{12}}{(\hbar\omega_{+}-P_{\bar{i}\sigma,1})(\hbar\omega_{+}-P_{\bar{i}\sigma,5})}]+\frac{D_{i\sigma,3}}{K_{i\sigma,5}}[U_{12}+\frac{t_{12}^2 U_{\bar{i}}}{(\hbar\omega_{+}-P_{\bar{i}\sigma,1})(\hbar\omega_{+}-P_{\bar{i}\sigma,2})}]
,\notag\\
C_{i\sigma,4}&=A_{i\sigma,4}\frac{C_{i\sigma,2}}{A_{i\sigma,2}}+A_{i\sigma,6}\frac{C_{i\sigma,3}}{A_{i\sigma,3}}+\frac{t_{12}^2 C_{i\sigma,1}}{K_{i\sigma,5}}(\frac{1}{\hbar\omega_{+}-P_{\bar{i}\sigma,3}}-\frac{1}{\hbar\omega_{+}-P_{\bar{i}\sigma,2}}-\frac{1}{\hbar\omega_{+}-P_{\bar{i}\sigma,5}}+\frac{1}{\hbar\omega_{+}-P_{\bar{i}\sigma,1}})\notag\\&+\frac{U_{12}}{K_{i\sigma,5}}(\frac{1}{\hbar\omega_{+}-P_{{i}\sigma,5}}-\frac{1}{\hbar\omega_{+}-P_{{i}\sigma,3}}-\frac{1}{\hbar\omega_{+}-P_{{i}\sigma,6}}+\frac{1}{\hbar\omega_{+}-P_{{i}\sigma,4}}).
\label{eq:coe_all}
\end{align}
By introducing the function for (numerical) integral 
\begin{equation}
    \phi^{'}(g)=-2\int f(\omega)\text{Im}[g(\omega)]
\end{equation}
and insert the coefficients in Eq. (\ref{eq:coe_all}), we arrive at
\begin{align}
    \langle \hat{n}_{i\bar{\sigma}}\hat{n}_{\bar{i}\bar{\sigma}}\hat{n}_{i\sigma}\rangle/\hbar&=\phi^{'}(B_{i\sigma,1})\langle\hat{n}_{i\bar{\sigma}}\hat{n}_{\bar{i}\bar{\sigma}}\rangle+\phi^{'}(C_{i\sigma,1})\langle\hat{n}_{i\bar{\sigma}}\hat{n}_{\bar{i}\sigma}\hat{n}_{\bar{i}\bar{\sigma}}\rangle,
    \label{eq:lin_4}
\end{align}

\begin{align}
   \langle \hat{n}_{i\bar{\sigma}}\hat{n}_{i\sigma}\rangle/\hbar&=\phi^{'}(A_{i\sigma,2})\langle\hat{n}_{i\bar{\sigma}}\rangle + \phi^{'}(B_{i\sigma,2})\langle\hat{n}_{i\bar{\sigma}}\hat{n}_{\bar{i}{\sigma}}\rangle + \phi^{'}(C_{i\sigma,2})\langle\hat{n}_{i\bar{\sigma}}\hat{n}_{\bar{i}\sigma}\hat{n}_{\bar{i}\bar{\sigma}}\rangle+\phi^{'}(D_{i\sigma,2})\langle\hat{n}_{i\bar{\sigma}}\hat{n}_{\bar{i}\bar{\sigma}}\rangle,
   \label{eq:lin_5}
\end{align}
\begin{align}
    \langle \hat{n}_{\bar{i}\bar{\sigma}}\hat{n}_{i\sigma}\rangle/\hbar&=\phi^{'}(A_{i\sigma,3})\langle\hat{n}_{\bar{i}\bar{\sigma}}\rangle + \phi^{'}(B_{i\sigma,3})\langle\hat{n}_{\bar{i}{\sigma}}\hat{n}_{\bar{i}\bar{\sigma}}\rangle + \phi^{'}(C_{i\sigma,3})\langle\hat{n}_{i\bar{\sigma}}\hat{n}_{\bar{i}\sigma}\hat{n}_{\bar{i}\bar{\sigma}}\rangle+ \phi^{'}(D_{i\sigma,3})\langle\hat{n}_{{i}\bar{\sigma}}\hat{n}_{\bar{i}\bar{\sigma}}\rangle,
    \label{eq:lin_6}
\end{align}

\begin{align}
    \langle \hat{n}_{i\sigma}\rangle/\hbar&=\phi^{'}(g_{i\sigma,0})+\phi^{'}(A_{i\sigma,4})\langle\hat{n}_{{i}\bar{\sigma}}\rangle + \phi^{'}(A_{i\sigma,5})\langle\hat{n}_{\bar{i}{\sigma}}\rangle + \phi^{'}(A_{i\sigma,6})\langle\hat{n}_{\bar{i}\bar{\sigma}}\rangle\notag\\&+
    \phi^{'}(B_{i\sigma,4})\langle\hat{n}_{{i}\bar{\sigma}}\hat{n}_{\bar{i}{\sigma}}\rangle + \phi^{'}(B_{i\sigma,5})\langle\hat{n}_{\bar{i}{\sigma}}\hat{n}_{\bar{i}\bar{\sigma}}\rangle + \phi^{'}(B_{i\sigma,6})\langle\hat{n}_{{i}\bar{\sigma}}\hat{n}_{\bar{i}\bar{\sigma}}\rangle +\phi^{'}(C_{i\sigma,4})\langle\hat{n}_{i\bar{\sigma}}\hat{n}_{\bar{i}\sigma}\hat{n}_{\bar{i}\bar{\sigma}}\rangle\label{eq:lin_7}.
\end{align}
Now we arrive at the linear system of density correlators given by Eqs. (\ref{eq:lin_1}, \ref{eq:lin_2}, \ref{eq:lin_3}, \ref{eq:lin_4}, \ref{eq:lin_5}, \ref{eq:lin_6},
\ref{eq:lin_7}).\\

In the following, we use $\hbar=1$ and arrive at a $14\times14$ linear system, which can be expressed as
\begin{equation}
    \boldsymbol{M}\boldsymbol{y}=\boldsymbol{b}.
    \label{eq:14by14}
\end{equation}
The vector of density correlators is given by
\begin{align}
    \boldsymbol{y}=&\bigg(\langle\hat{n}_{1\uparrow}\rangle,\langle\hat{n}_{1\downarrow}\rangle,\langle\hat{n}_{2\uparrow}\rangle,\langle\hat{n}_{2\downarrow}\rangle,\langle\hat{n}_{1\uparrow}\hat{n}_{1\downarrow}\rangle,\langle\hat{n}_{1\uparrow}\hat{n}_{2\uparrow}\rangle,\langle\hat{n}_{1\uparrow}\hat{n}_{2\downarrow}\rangle,\langle\hat{n}_{1\downarrow}\hat{n}_{2\uparrow}\rangle,\langle\hat{n}_{1\downarrow}\hat{n}_{2\downarrow}\rangle,\langle\hat{n}_{2\uparrow}\hat{n}_{2\downarrow}\rangle,\notag\\&\langle\hat{n}_{1\uparrow}\hat{n}_{1\downarrow}\hat{n}_{2\uparrow}\rangle,\langle\hat{n}_{1\uparrow}\hat{n}_{1\downarrow}\hat{n}_{2\downarrow}\rangle,\langle\hat{n}_{2\uparrow}\hat{n}_{2\downarrow}\hat{n}_{1\uparrow}\rangle,\langle\hat{n}_{2\uparrow}\hat{n}_{2\downarrow}\hat{n}_{1\downarrow}\rangle\bigg)^T.
    \label{eq:unknown}
\end{align}
The vector $\boldsymbol{b}$ is given by
\begin{equation}
    \boldsymbol{b}=\bigg(\phi^{'}(g_{1\uparrow,0}), \phi^{'}(g_{1\downarrow,0}),\phi^{'}(g_{2\uparrow,0}),\phi^{'}(g_{2\downarrow,0}),0,0,0,0,0,0,0,0,0,0\bigg)^T.\label{eq:y}
\end{equation}
The nonzero elements of the $14\times14$ matrix $\boldsymbol{M}$, i.e., ${M}_{i,j}$ with $i$ ($j$) being the row (column) index, are explicitly given in the next section based on a symmetry- preserving treatment of the EOM approach.

\subsection{Symmetry-preserving EOM treatment}
When constructing the $14\times14$ system of density correlators in
Eq. (\ref{eq:14by14}), a given correlator can generally be obtained through more than one
formally equivalent EOM route. For example,
$\langle \hat n_{1\uparrow}\hat n_{2\uparrow}\rangle$ can be obtained by
choosing either $i=1,\sigma=\uparrow$ or $i=2,\sigma=\uparrow$ in
Eq. (\ref{eq:lin_3}). We denote the two results as
\begin{equation}
\left\langle\hat n_{1\uparrow}\hat n_{2\uparrow}\right\rangle_
{1\uparrow\rightarrow2\uparrow},
\qquad
\left\langle\hat n_{1\uparrow}\hat n_{2\uparrow}\right\rangle_
{2\uparrow\rightarrow1\uparrow}.
\end{equation}
Since the density operators commute, the two routes yield exactly the same
correlator in the untruncated EOM hierarchy,
\begin{equation}
\left\langle\hat n_{1\uparrow}\hat n_{2\uparrow}\right\rangle_
{1\uparrow\rightarrow2\uparrow}
=
\left\langle\hat n_{1\uparrow}\hat n_{2\uparrow}\right\rangle_
{2\uparrow\rightarrow1\uparrow}.
\end{equation}
A finite-order EOM closure, however, can introduce a spurious route dependence
through the truncated higher-order Green's functions. To remove this artificial
asymmetry, we retain the equal-weight combination of the two equivalent routes,
\begin{equation}
\left\langle\hat n_{1\uparrow}\hat n_{2\uparrow}\right\rangle
=
\frac{1}{2}
\left[
\left\langle\hat n_{1\uparrow}\hat n_{2\uparrow}\right\rangle_
{1\uparrow\rightarrow2\uparrow}
+
\left\langle\hat n_{1\uparrow}\hat n_{2\uparrow}\right\rangle_
{2\uparrow\rightarrow1\uparrow}
\right].\label{eq:avg}
\end{equation}
The same symmetrization is applied to all correlators possessing equivalent
EOM routes. This symmetry-preserving closure
removes the artificial route dependence introduced by the EOM truncation
without imposing additional physical symmetries. \jacob{Genuine} asymmetries arising
from unequal gate energies or Coulomb interactions, as well as the spin
asymmetry induced by $M_z$, are fully retained. Accordingly, all density
correlators entering Eqs. (\ref{eq:lin_1}, \ref{eq:lin_2}, \ref{eq:lin_3}, \ref{eq:lin_4}, \ref{eq:lin_5}, \ref{eq:lin_6},
\ref{eq:lin_7}) that admit two equivalent EOM routes are
evaluated from their equal-weight combination as in Eq. (\ref{eq:avg}). Related
symmetrization procedures have also been employed in EOM treatments to
preserve particle-hole symmetry \cite{EOMsym}.

Based on the above symmetry-preserving treatment, we obtain the explicit nonzero matrix elements in $\boldsymbol{M}$ as:
\begin{align}
M_{1,1}&=1,M_{1,2}=-\phi^{'}(A_{1\uparrow,4}), M_{1,3}=-\phi^{'}(A_{1\uparrow,5}),M_{1,4}=-\phi^{'}(A_{1\uparrow,6}),\notag\\ M_{1,8}&=-\phi^{'}(B_{1\uparrow,4}), 
M_{1,9}=-\phi^{'}(B_{1\uparrow,6}), M_{1,10}=-\phi^{'}(B_{1\uparrow,5}), M_{1,14}=-\phi^{'}(C_{1\uparrow,4}),\\
M_{2,1}&=-\phi^{'}(A_{1\downarrow,4}), M_{2,2}=1, M_{2,3}=-\phi^{'}(A_{1\downarrow,6}), M_{2,4}=-\phi^{'}(A_{1\downarrow,5}),\notag\\ 
M_{2,6}&=-\phi^{'}(B_{1\downarrow,6}),
M_{2,7}=-\phi^{'}(B_{1\downarrow,4}), M_{2,10}=-\phi^{'}(B_{1\downarrow,5}), M_{2,13}=-\phi^{'}(C_{1\downarrow,4}),\\
M_{3,1}&=-\phi^{'}(A_{2\uparrow,5}), M_{3,2}=-\phi^{'}(A_{2\uparrow,6}), M_{3,3}=1, M_{3,4}=-\phi^{'}(A_{2\uparrow,4}), \notag\\M_{3,7}&=-\phi^{'}(B_{2\uparrow,4}), M_{3,5}=-\phi^{'}(B_{2\uparrow,5}), 
M_{3,9}=-\phi^{'}(B_{2\uparrow,6}),
M_{3,12}=-\phi^{'}(C_{2\uparrow,4}),\\
M_{4,1}&=-\phi^{'}(A_{2\downarrow,6}), M_{4,2}=-\phi^{'}(A_{2\downarrow,5}), M_{4,3}=-\phi^{'}(A_{2\downarrow,4}), M_{4,4}=1, \notag\\
M_{4,5}&=-\phi^{'}(B_{2\downarrow,5}),
M_{4,6}=-\phi^{'}(B_{2\downarrow,6}),
M_{4,8}=-\phi^{'}(B_{2\downarrow,4}), M_{4,11}=-\phi^{'}(C_{2\downarrow,4}),\\
M_{5,2}&=-\phi^{'}(A_{1\uparrow,2})/2, \quad M_{5,5}=1, \quad M_{5,8}=-\phi^{'}(B_{1\uparrow,2})/2, \quad M_{5,9}=-\phi^{'}(D_{1\uparrow,2})/2,\quad M_{5,14}=-\phi^{'}(C_{1\uparrow,2})/2,\notag\\M_{5,1}&=-\phi^{'}(A_{1\downarrow,2})/2, \quad M_{5,7}=-\phi^{'}(B_{1\downarrow,2})/2, \quad M_{5,6}=-\phi^{'}(D_{1\downarrow,2})/2,\quad M_{5,13}=-\phi^{'}(C_{1\downarrow,2})/2,\\
M_{6,1}&=-\phi(p_{2\uparrow,5})/2,\quad M_{6,3}=-\phi(p_{1\uparrow,5})/2, \quad M_{6,5}=\phi(p_{2\uparrow,5})/2-\phi(p_{2\uparrow,6})/2, \quad M_{6,6}=1, \quad M_{6,7}=\phi(p_{2\uparrow,5})/2-\phi(p_{2\uparrow,3})/2, \notag\\  M_{6,8}&=\phi(p_{1\uparrow,5})/2-\phi(p_{1\uparrow,3})/2,  \quad M_{6,10}=\phi(p_{1\uparrow,5})/2-\phi(p_{1\uparrow,6})/2,\notag\\ M_{6,12}&=-\phi(p_{2\uparrow,5})/2+\phi(p_{2\uparrow,3})/2+\phi(p_{2\uparrow,6})/2-\phi(p_{2\uparrow,4})/2,\quad M_{6,14}=-\phi(p_{1\uparrow,5})/2+\phi(p_{1\uparrow,3})/2+\phi(p_{1\uparrow,6})/2-\phi(p_{1\uparrow,4})/2,\\
M_{7,1}&=-\phi^{'}(A_{2\downarrow,3})/2,\quad 
M_{7,4}=-\phi^{'}(A_{1\uparrow,3})/2, \quad M_{7,5}=-\phi^{'}(B_{2\downarrow,3})/2,\quad M_{7,6}=-\phi^{'}(D_{2\downarrow,3})/2,\quad M_{7,7}=1,\notag\\ M_{7,9}&=-\phi^{'}(D_{1\uparrow,3})/2,\quad M_{7,10}=-\phi^{'}(B_{1\uparrow,3})/2,\quad M_{7,11}=-\phi^{'}(C_{2\downarrow,3})/2,\quad M_{7,14}=-\phi^{'}(C_{1\uparrow,3})/2,\\
M_{8,2}&=-\phi^{'}(A_{2\uparrow,3})/2,\quad M_{8,3}=-\phi^{'}(A_{1\downarrow,3})/2,\quad M_{8,5}=-\phi^{'}(B_{2\uparrow,3})/2,\quad M_{8,6}=-\phi^{'}(D_{1\downarrow,3})/2,\quad M_{8,8}=1,\notag\\M_{8,9}&=-\phi^{'}(D_{2\uparrow,3})/2, \quad M_{8,10}=-\phi^{'}(B_{1\downarrow,3})/2, \quad M_{8,12}=-\phi^{'}(C_{2\uparrow,3})/2,\quad M_{8,13}=-\phi^{'}(C_{1\downarrow,3})/2,\\
M_{9,2}&=-\phi(p_{2\downarrow,5})/2,\quad M_{9,4}=-\phi(p_{1\downarrow,5})/2,\quad M_{9,5}=\phi(p_{2\downarrow,5})/2-\phi(p_{2\downarrow,6})/2,\quad M_{9,7}=\phi(p_{1\downarrow,5})/2-\phi(p_{1\downarrow,3})/2, \notag\\ M_{9,8}&=\phi(p_{2\downarrow,5})/2-\phi(p_{2\downarrow,3})/2,\quad M_{9,9}=1,\quad M_{9,10} =\phi(p_{1\downarrow,5})/2-\phi(p_{1\downarrow,6})/2,\notag\\M_{9,11}&=-\phi(p_{2\downarrow,5})/2+\phi(p_{2\downarrow,3})/2+\phi(p_{2\downarrow,6})/2-\phi(p_{2\downarrow,4})/2,\quad M_{9,13}=-\phi(p_{1\downarrow,5})/2+\phi(p_{1\downarrow,3})/2+\phi(p_{1\downarrow,6})/2-\phi(p_{1\downarrow,4})/2,\\
M_{10,3}&=-\phi^{'}(A_{2\downarrow,2})/2,\quad M_{10,6}=-\phi^{'}(D_{2\downarrow,2})/2,\quad M_{10,8}=-\phi^{'}(B_{2\downarrow,2})/2, \quad M_{10,10}=1,\quad M_{10,11}=-\phi^{'}(C_{2\downarrow,2})/2,\notag\\M_{10,4}&=-\phi^{'}(A_{2\uparrow,2})/2,\quad M_{10,7}=-\phi^{'}(B_{2\uparrow,2})/2,\quad M_{10,9}=-\phi^{'}(D_{2\uparrow,2})/2, \quad M_{10,12}=-\phi^{'}(C_{2\uparrow,2})/2,\\
M_{11,5}&=-\phi(p_{2\uparrow,6})/2,\quad M_{11,8}=-\phi(p_{1\uparrow,3})/2,\quad M_{11,11}=1, \quad M_{11,12}=\phi(p_{2\uparrow,6})/2-\phi(p_{2\uparrow,4})/2,\notag\\M_{11,14}&=\phi(p_{1\uparrow,3})/2-\phi(p_{1\uparrow,4})/2,\\
M_{12,5}&=-\phi(p_{2\downarrow,6})/2,\quad M_{12,7}=-\phi(p_{1\downarrow,3})/2,\quad M_{12,11}=\phi(p_{2\downarrow,6})/2-\phi(p_{2\downarrow,4})/2,\quad M_{12,12}=1,\notag\\ M_{12,13}&=\phi(p_{1\downarrow,3})/2-\phi(p_{1\downarrow,4})/2,\\M_{13,7}&=-\phi(p_{2\uparrow,3})/2,\quad 
M_{13,10}=-\phi(p_{1\uparrow,6})/2,\quad \quad M_{13,12}=\phi(p_{2\uparrow,3})/2-\phi(p_{2\uparrow,4})/2,\quad M_{13,13}=1, \notag\\M_{13,14}&=\phi(p_{1\uparrow,6})/2-\phi(p_{1\uparrow,4})/2,\\
M_{14,8}&=-\phi(p_{2\downarrow,3})/2,\quad M_{14,10}=-\phi(p_{1\downarrow,6})/2, \quad M_{14,11}=\phi(p_{2\downarrow,3})/2-\phi(p_{2\downarrow,4})/2,\quad M_{14,14}=1,\notag\\M_{14,13}&=\phi(p_{1\downarrow,6})/2-\phi(p_{1\downarrow,4})/2.
\end{align}
Given the explicit expression of $\boldsymbol{M}$ above, the one-, two- and three-body density correlators listed in Eq. (\ref{eq:unknown}) can be obtained by solving Eq. (\ref{eq:14by14}) numerically.

\subsection{Four-body density correlator}
In the previous section, the one-,two-, three-body densities correlators given in Eq. (\ref{eq:unknown}) can be solved. Now we turn to the highest-order four-body density correlator.\\

As shown in Eq. (\ref{eq:Dnnn1_new}), we have the highest-order GF
    \begin{align}
    \langle\langle \hat{n}_{i\bar{\sigma}}\hat{n}_{\bar{i}\sigma}\hat{n}_{\bar{i}\bar{\sigma}}\hat{d}_{i\sigma}:\hat{d}_{i\sigma}^\dag\rangle\rangle_\omega&=\frac{\hbar\langle\hat{n}_{i\bar{\sigma}}\hat{n}_{\bar{i}\sigma}\hat{n}_{\bar{i}\bar{\sigma}}\rangle}{\hbar\omega_+ - P_{i\sigma,4}}.
    \label{eq:highest}
\end{align}
We then insert Eq. (\ref{eq:highest}) into Eq. (\ref{eq:D_to_n_new}) to get the four-body density correlators in terms of the three-body ones. Note the four-body correlator can be evaluated through the four
equivalent routes $(i,\sigma)=(1,\uparrow),(1,\downarrow),
(2,\uparrow),(2,\downarrow)$, yielding
\begin{align}
\left\langle
\hat n_{1\uparrow}\hat n_{1\downarrow}
\hat n_{2\uparrow}\hat n_{2\downarrow}
\right\rangle
=\frac{1}{4}\Big[
&
\phi'(p_{1\uparrow,4})
\left\langle
\hat n_{2\uparrow}\hat n_{2\downarrow}\hat n_{1\downarrow}
\right\rangle
+
\phi'(p_{1\downarrow,4})
\left\langle
\hat n_{2\uparrow}\hat n_{2\downarrow}\hat n_{1\uparrow}
\right\rangle
\nonumber\\
&+
\phi'(p_{2\uparrow,4})
\left\langle
\hat n_{1\uparrow}\hat n_{1\downarrow}\hat n_{2\downarrow}
\right\rangle
+
\phi'(p_{2\downarrow,4})
\left\langle
\hat n_{1\uparrow}\hat n_{1\downarrow}\hat n_{2\uparrow}
\right\rangle
\Big].
\end{align}

\section{Many-body state probabilities: 16 Fock states}
Unlike the standard mean-field/NEGF approaches often only give the one-body density correlator $\langle\hat{n}_{i\sigma}\rangle$ well, our EOM approach has the full access of all density correlators up to fourth order, based on which we can reconstruct the probabilities of the many-body configurations of the double quantum dot. We employ the occupation-number basis $|ABCD\rangle$ with
$A,B,C,D\in\{0,1\}$,
where the occupations correspond to the spin orbitals
$(1\uparrow,1\downarrow,2\uparrow,2\downarrow)$ with $i=1,2$ denotes the QD site and $\sigma=\uparrow,\downarrow$ being the spin. The probability of the many-body state $|ABCD\rangle$ is

\begin{align}
P_{ABCD}
=
\langle
\hat n_{1\uparrow}^{A}
(1-\hat n_{1\uparrow})^{1-A}
\hat n_{1\downarrow}^{B}
(1-\hat n_{1\downarrow})^{1-B}
\hat n_{2\uparrow}^{C}
(1-\hat n_{2\uparrow})^{1-C}
\hat n_{2\downarrow}^{D}
(1-\hat n_{2\downarrow})^{1-D}
\rangle .
\label{eq:P_ABCD}
\end{align}
We then have the 16 spin-resolved Fock states:
\begin{align}
|0,0\rangle: P_{0000}
=&\,1-\langle \hat n_{1\uparrow}\rangle-\langle \hat n_{1\downarrow}\rangle
-\langle \hat n_{2\uparrow}\rangle-\langle \hat n_{2\downarrow}\rangle+\langle \hat n_{1\uparrow}\hat n_{1\downarrow}\rangle
+\langle \hat n_{1\uparrow}\hat n_{2\uparrow}\rangle
+\langle \hat n_{1\uparrow}\hat n_{2\downarrow}\rangle
+\langle \hat n_{1\downarrow}\hat n_{2\uparrow}\rangle
+\langle \hat n_{1\downarrow}\hat n_{2\downarrow}\rangle
+\langle \hat n_{2\uparrow}\hat n_{2\downarrow}\rangle
\nonumber\\
&-\langle \hat n_{1\uparrow}\hat n_{1\downarrow}\hat n_{2\uparrow}\rangle
-\langle \hat n_{1\uparrow}\hat n_{1\downarrow}\hat n_{2\downarrow}\rangle
-\langle \hat n_{1\uparrow}\hat n_{2\uparrow}\hat n_{2\downarrow}\rangle
-\langle \hat n_{1\downarrow}\hat n_{2\uparrow}\hat n_{2\downarrow}\rangle
+\langle \hat n_{1\uparrow}\hat n_{1\downarrow}\hat n_{2\uparrow}\hat n_{2\downarrow}\rangle ,
\end{align}

\begin{equation}
|\uparrow,0\rangle: P_{1000}
=\langle \hat n_{1\uparrow}\rangle
-\langle \hat n_{1\uparrow}\hat n_{1\downarrow}\rangle
-\langle \hat n_{1\uparrow}\hat n_{2\uparrow}\rangle
-\langle \hat n_{1\uparrow}\hat n_{2\downarrow}\rangle
+\langle \hat n_{1\uparrow}\hat n_{1\downarrow}\hat n_{2\uparrow}\rangle
+\langle \hat n_{1\uparrow}\hat n_{1\downarrow}\hat n_{2\downarrow}\rangle
+\langle \hat n_{1\uparrow}\hat n_{2\uparrow}\hat n_{2\downarrow}\rangle
-\langle \hat n_{1\uparrow}\hat n_{1\downarrow}\hat n_{2\uparrow}\hat n_{2\downarrow}\rangle ,
\end{equation}

\begin{equation}
|\downarrow,0\rangle: P_{0100}
=\langle \hat n_{1\downarrow}\rangle
-\langle \hat n_{1\uparrow}\hat n_{1\downarrow}\rangle
-\langle \hat n_{1\downarrow}\hat n_{2\uparrow}\rangle
-\langle \hat n_{1\downarrow}\hat n_{2\downarrow}\rangle
+\langle \hat n_{1\uparrow}\hat n_{1\downarrow}\hat n_{2\uparrow}\rangle
+\langle \hat n_{1\uparrow}\hat n_{1\downarrow}\hat n_{2\downarrow}\rangle
+\langle \hat n_{1\downarrow}\hat n_{2\uparrow}\hat n_{2\downarrow}\rangle
-\langle \hat n_{1\uparrow}\hat n_{1\downarrow}\hat n_{2\uparrow}\hat n_{2\downarrow}\rangle ,
\end{equation}

\begin{equation}
|0,\uparrow\rangle: P_{0010}
=\langle \hat n_{2\uparrow}\rangle
-\langle \hat n_{1\uparrow}\hat n_{2\uparrow}\rangle
-\langle \hat n_{1\downarrow}\hat n_{2\uparrow}\rangle
-\langle \hat n_{2\uparrow}\hat n_{2\downarrow}\rangle
+\langle \hat n_{1\uparrow}\hat n_{1\downarrow}\hat n_{2\uparrow}\rangle
+\langle \hat n_{1\uparrow}\hat n_{2\uparrow}\hat n_{2\downarrow}\rangle
+\langle \hat n_{1\downarrow}\hat n_{2\uparrow}\hat n_{2\downarrow}\rangle
-\langle \hat n_{1\uparrow}\hat n_{1\downarrow}\hat n_{2\uparrow}\hat n_{2\downarrow}\rangle ,
\end{equation}

\begin{equation}
|0,\downarrow\rangle: P_{0001}
=\langle \hat n_{2\downarrow}\rangle
-\langle \hat n_{1\uparrow}\hat n_{2\downarrow}\rangle
-\langle \hat n_{1\downarrow}\hat n_{2\downarrow}\rangle
-\langle \hat n_{2\uparrow}\hat n_{2\downarrow}\rangle
+\langle \hat n_{1\uparrow}\hat n_{1\downarrow}\hat n_{2\downarrow}\rangle
+\langle \hat n_{1\uparrow}\hat n_{2\uparrow}\hat n_{2\downarrow}\rangle
+\langle \hat n_{1\downarrow}\hat n_{2\uparrow}\hat n_{2\downarrow}\rangle
-\langle \hat n_{1\uparrow}\hat n_{1\downarrow}\hat n_{2\uparrow}\hat n_{2\downarrow}\rangle ,
\end{equation}

\begin{equation}
|\uparrow\downarrow,0\rangle:P_{1100}
=\langle \hat n_{1\uparrow}\hat n_{1\downarrow}\rangle
-\langle \hat n_{1\uparrow}\hat n_{1\downarrow}\hat n_{2\uparrow}\rangle
-\langle \hat n_{1\uparrow}\hat n_{1\downarrow}\hat n_{2\downarrow}\rangle
+\langle \hat n_{1\uparrow}\hat n_{1\downarrow}\hat n_{2\uparrow}\hat n_{2\downarrow}\rangle ,
\end{equation}

\begin{equation}
|\uparrow,\uparrow\rangle:P_{1010}
=\langle \hat n_{1\uparrow}\hat n_{2\uparrow}\rangle
-\langle \hat n_{1\uparrow}\hat n_{1\downarrow}\hat n_{2\uparrow}\rangle
-\langle \hat n_{1\uparrow}\hat n_{2\uparrow}\hat n_{2\downarrow}\rangle
+\langle \hat n_{1\uparrow}\hat n_{1\downarrow}\hat n_{2\uparrow}\hat n_{2\downarrow}\rangle ,
\end{equation}

\begin{equation}
|\uparrow,\downarrow\rangle:P_{1001}
=\langle \hat n_{1\uparrow}\hat n_{2\downarrow}\rangle
-\langle \hat n_{1\uparrow}\hat n_{1\downarrow}\hat n_{2\downarrow}\rangle
-\langle \hat n_{1\uparrow}\hat n_{2\uparrow}\hat n_{2\downarrow}\rangle
+\langle \hat n_{1\uparrow}\hat n_{1\downarrow}\hat n_{2\uparrow}\hat n_{2\downarrow}\rangle ,
\end{equation}

\begin{equation}
|\downarrow,\uparrow\rangle:P_{0110}
=\langle \hat n_{1\downarrow}\hat n_{2\uparrow}\rangle
-\langle \hat n_{1\uparrow}\hat n_{1\downarrow}\hat n_{2\uparrow}\rangle
-\langle \hat n_{1\downarrow}\hat n_{2\uparrow}\hat n_{2\downarrow}\rangle
+\langle \hat n_{1\uparrow}\hat n_{1\downarrow}\hat n_{2\uparrow}\hat n_{2\downarrow}\rangle ,
\end{equation}

\begin{equation}
|\downarrow,\downarrow\rangle: P_{0101}
=\langle \hat n_{1\downarrow}\hat n_{2\downarrow}\rangle
-\langle \hat n_{1\uparrow}\hat n_{1\downarrow}\hat n_{2\downarrow}\rangle
-\langle \hat n_{1\downarrow}\hat n_{2\uparrow}\hat n_{2\downarrow}\rangle
+\langle \hat n_{1\uparrow}\hat n_{1\downarrow}\hat n_{2\uparrow}\hat n_{2\downarrow}\rangle ,
\end{equation}

\begin{equation}
|0,\uparrow\uparrow\rangle:P_{0011}
=\langle \hat n_{2\uparrow}\hat n_{2\downarrow}\rangle
-\langle \hat n_{1\uparrow}\hat n_{2\uparrow}\hat n_{2\downarrow}\rangle
-\langle \hat n_{1\downarrow}\hat n_{2\uparrow}\hat n_{2\downarrow}\rangle
+\langle \hat n_{1\uparrow}\hat n_{1\downarrow}\hat n_{2\uparrow}\hat n_{2\downarrow}\rangle ,
\end{equation}

\begin{equation}
|\uparrow\downarrow,\uparrow\rangle:P_{1110}
=\langle \hat n_{1\uparrow}\hat n_{1\downarrow}\hat n_{2\uparrow}\rangle
-\langle \hat n_{1\uparrow}\hat n_{1\downarrow}\hat n_{2\uparrow}\hat n_{2\downarrow}\rangle ,
\end{equation}

\begin{equation}
|\uparrow\downarrow,\downarrow\rangle:P_{1101}
=\langle \hat n_{1\uparrow}\hat n_{1\downarrow}\hat n_{2\downarrow}\rangle
-\langle \hat n_{1\uparrow}\hat n_{1\downarrow}\hat n_{2\uparrow}\hat n_{2\downarrow}\rangle ,
\end{equation}

\begin{equation}
|\uparrow,\uparrow\downarrow\rangle: P_{1011}
=\langle \hat n_{1\uparrow}\hat n_{2\uparrow}\hat n_{2\downarrow}\rangle
-\langle \hat n_{1\uparrow}\hat n_{1\downarrow}\hat n_{2\uparrow}\hat n_{2\downarrow}\rangle ,
\end{equation}

\begin{equation}
|\downarrow,\uparrow\downarrow\rangle: P_{0111}
=\langle \hat n_{1\downarrow}\hat n_{2\uparrow}\hat n_{2\downarrow}\rangle
-\langle \hat n_{1\uparrow}\hat n_{1\downarrow}\hat n_{2\uparrow}\hat n_{2\downarrow}\rangle ,
\end{equation}

\begin{equation}
|\uparrow\downarrow,\uparrow\downarrow\rangle:P_{1111}
=\langle \hat n_{1\uparrow}\hat n_{1\downarrow}\hat n_{2\uparrow}\hat n_{2\downarrow}\rangle .
\end{equation}



    

\section{Consistency assessment of the EOM truncation}
\label{sec:EOM_consistency}

The EOM hierarchy employed in this work is truncated in the interaction-dominated regime, where the dot--reservoir coupling and interdot tunneling are smaller energy scales than the Coulomb interactions. Because such a finite closure does not automatically guarantee all properties of the exact hierarchy, we performed several numerical and symmetry-based tests of the resulting solutions.\\

Using the occupation-state notation \(P_{ABCD}\) introduced in Eq. (\ref{eq:P_ABCD}), we define the left- and right-localized single-electron probabilities as
\begin{equation}
P_L=P_{1000}+P_{0100},
\qquad
P_R=P_{0010}+P_{0001},
\label{eq:PL_PR_definition}
\end{equation}
where \(P_{1000}\), \(P_{0100}\), \(P_{0010}\), and \(P_{0001}\) correspond to the single-electron occupation states \(1\uparrow\), \(1\downarrow\), \(2\uparrow\), and \(2\downarrow\), respectively. Physically, \(P_L\) is the probability that the DQD contains exactly one electron localized on dot 1, irrespective of its spin, whereas \(P_R\) is the corresponding probability for localization on dot 2. These quantities therefore give the statistical weights of the left- and right-localized states forming the charge-qubit subspace. Their sum,
\begin{equation}
P_{N=1}=P_L+P_R,
\label{eq:PN1_definition}
\end{equation}
is the total probability that the DQD occupies the single-electron sector. Accordingly, \(P_L-P_R\) measures the charge polarization within this sector, and a sign reversal of \(P_L-P_R\) indicates switching of the dominant charge-qubit occupation from one dot to the other. The complementary probability $1-P_{N=1}$ quantifies leakage into the empty and higher-charge sectors.\\

\subsection{ Preservation of probability positivity and normalization}
First, we examined the complete occupation-state probability distribution reconstructed from the one- through four-body density correlators. Throughout the parameter regimes considered in the main text, all 16 occupation-state probabilities satisfy
\begin{equation}
P_{ABCD}\geq 0,
\qquad
\sum_{A=0}^{1}\sum_{B=0}^{1}
\sum_{C=0}^{1}\sum_{D=0}^{1}P_{ABCD}=1
\label{eq:probability_consistency}
\end{equation}
within numerical precision. The EOM solution therefore produces a positive and normalized probability distribution over the complete four-mode Fock space.\\

\subsection{Preservation of spin-reversal symmetry}
Second, the calculation satisfies the spin-reversal relations imposed by the Hamiltonian. Because the reservoirs are spin independent and the magnetic field enters only through the uniform Zeeman term, reversing \(M_z\) is equivalent to interchanging the spin labels. The four single-electron probabilities consequently obey
\begin{equation}
P_{1000}(M_z)=P_{0100}(-M_z),
\qquad
P_{0010}(M_z)=P_{0001}(-M_z).
\label{eq:spin_reversal}
\end{equation}
It follows from Eq.~\eqref{eq:PL_PR_definition} that the spin-summed left- and right-dot probabilities are even functions of \(M_z\):
\begin{equation}
P_L(M_z)=P_L(-M_z),
\qquad
P_R(M_z)=P_R(-M_z).
\label{eq:charge_even}
\end{equation}
The magnetic charge reconfiguration is therefore controlled by \(\lvert M_z\rvert\), whereas the sign of \(M_z\) selects the dominant spin-resolved occupation state.\\

\subsection{Preservation of left-right symmetry}
Third, we verified the expected left--right symmetry of the DQD. When the two dots and their reservoir couplings are identical, the calculation recovers
\begin{equation}
P_{1000}(M_z)=P_{0010}(M_z),
\qquad
P_{0100}(M_z)=P_{0001}(M_z),
\label{eq:state_LR_symmetry}
\end{equation}
and hence
\begin{equation}
P_L(M_z)=P_R(M_z)
\label{eq:LR_symmetry}
\end{equation}
for every \(M_z\). In particular, for \(v_1=v_2\), \(U_1=U_2\), and otherwise identical dot and reservoir parameters, the uniform Zeeman splitting does not generate a left--right charge polarization. The finite response obtained at \(v_1=v_2\) when \(U_1\neq U_2\) therefore reflects the physical asymmetry of the many-body addition spectrum rather than an artificial dot asymmetry introduced by the EOM construction.\\

\subsection{Summary of the tests}
Finally, a finite EOM closure can make a given density correlator weakly dependent on the formally equivalent route through which its equation is generated. We remove this truncation-induced route dependence by averaging all equivalent EOM routes or representations with equal weights. This symmetry-preserving construction restores the equivalence of correlators related by permutations of commuting density operators while retaining the physical asymmetries associated with \(v_1\neq v_2\), \(U_1\neq U_2\), and finite \(M_z\).\\

The simultaneous preservation of probability positivity and normalization, spin-reversal symmetry, left--right symmetry in the identical-dot limit, and equivalence among different EOM routes provides internal consistency tests of the truncated EOM approach in the parameter regime considered here.

\section{Robustness of magnetic switching at reduced interdot interaction}

In the main text, we use $U_1=2$, $U_2=4$, and $U_{12}=3.5$ to
clearly expose the interaction-driven switching regime. While
$U_{12}>\sqrt{U_1U_2}$ lies beyond the constraint associated with a
conventional two-dot capacitance parametrization, this constraint is not
general to an effective interacting DQD Hamiltonian, for which strong
interdot-interaction regimes with $U_{12}>U_i$ have been considered \cite{nishino2016exact,sobrinoprb24,coll2021density}. Here we demonstrate that the predicted switching does not rely
on this strong-interaction choice.\\

Figure \ref{fig:switch1} shows the corresponding results for $U_{12}=2.5$, with all
other interaction parameters unchanged. This value satisfies the
conventional capacitive condition $U_{12}<\sqrt{U_1U_2}=\sqrt{8}\simeq2.83$.
As shown in Fig. \ref{fig:switch1} (a), a pronounced field-induced redistribution
persists over an extended region of gate space while
$P_{N=1}\geq0.95$. More importantly, Fig. \ref{fig:switch1} (b) reveals a broad
switching region near $v_1\simeq v_2$. At the representative point
$v_1=v_2=-1.1$, the left- and right-dot probabilities cross as $M_z$
is varied [Fig. \ref{fig:switch1} (c)], while their sum remains close to unity.
The accompanying reorganization of the spin-resolved single-electron
states is shown in Fig. \ref{fig:switch1} (d).\\

The persistence of charge-polarization reversal at $U_{12}=2.5$
demonstrates that Zeeman-driven switching is not contingent on the
larger $U_{12}=3.5$ used in the main text and survives within the
conventional capacitive regime. This provides an independent robustness
check of the interaction-driven switching mechanism.

\begin{figure}
\includegraphics[width=0.6\linewidth]{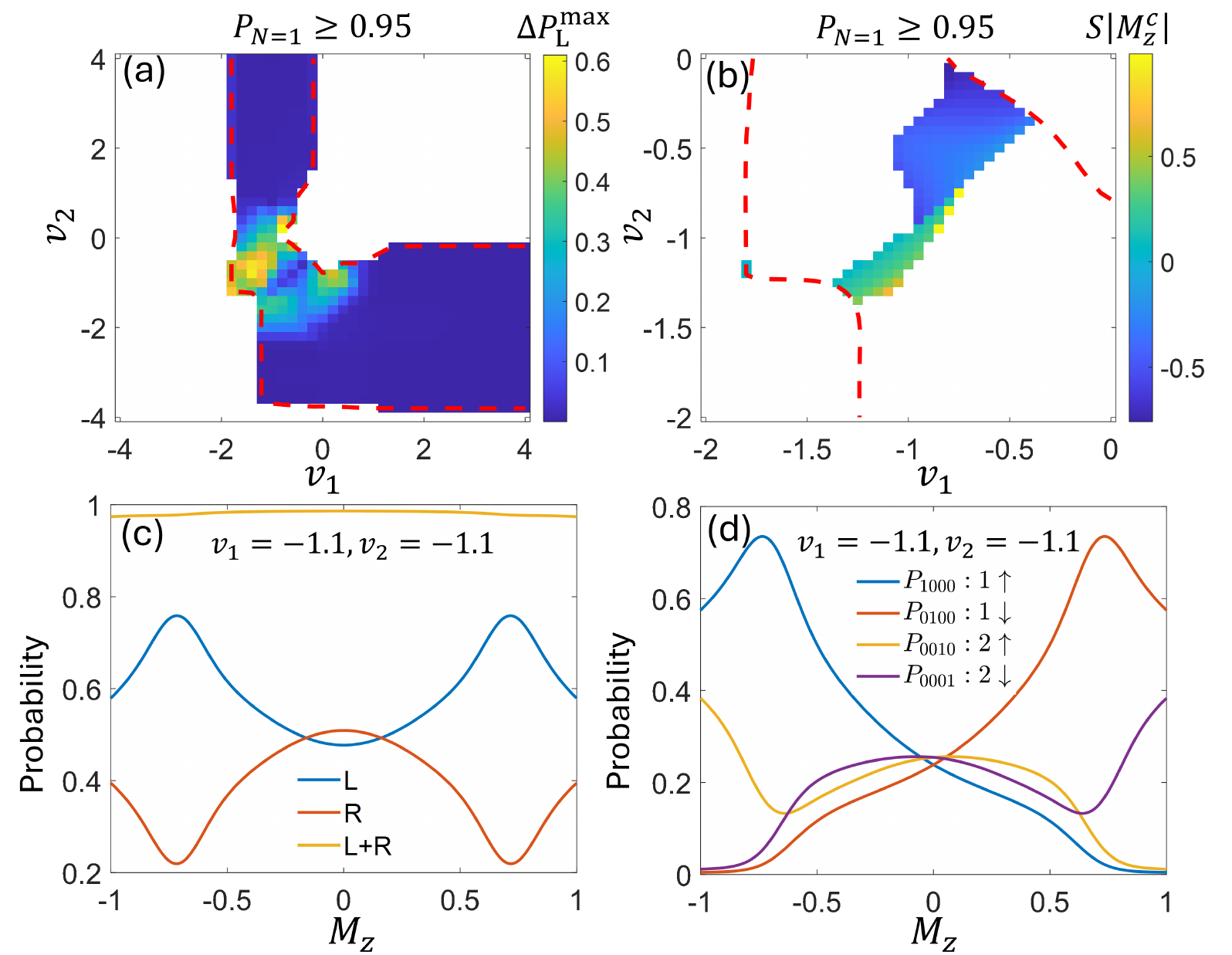}
	\caption{Magnetic charge switching for $U_1\neq U_2$.
(a) Maximum charge redistribution $\Delta P_L^{\max}$.
(b) Signed switching threshold $S|M_z^c|$, where
$P_L(M_z^c)=P_R(M_z^c)$ and $S>0$ ($S<0$) denotes right-to-left
(left-to-right) switching.
(c) Left, right, and total single-electron probabilities and
(d) spin-resolved probabilities versus $M_z$ at $v_1=v_2=-1.1$.
{The red dashed line in (a,b) denotes the boundary $P_{N=1}=0.95$ throughout $|M_z|\leq1$ and the white regions outside the red dashed line do not satisfy the $P_{N=1}\geq0.95$ criterion.}
Here $U_1=2$, $U_2=4$ and $U_{12}=2.5$, satisfying $U_{12}<\sqrt{U_1U_2}$; other parameters are as in
Fig. 3 in the main text.} 
    \label{fig:switch1}
\end{figure}

\section{Derivation of Eq. (\ref{eq:dev}) and Eq. (\ref{eq:D_to_n_new})}
The lesser Green function is defined as
\begin{equation}
    G_{i\sigma}^{<}(t,t^{'})=i\langle d_{i\sigma}^\dag(t^{'})d_{i\sigma}(t)\rangle.
\end{equation}
Then we set $t=t^{'}$ and have
\begin{equation}
    \langle n_{i\sigma}\rangle=\langle d_{i\sigma}^\dag(t)d_{i\sigma}(t)\rangle=-iG_{i\sigma}^{<}(t,t).
    \label{eq:ntt}
\end{equation}
Perform the Fourier transform, we have
\begin{equation}
    G_{i\sigma}^{<}(t,t)=\int\frac{d\omega}{2\pi}G_{i\sigma}^{<}(\omega).
    \label{eq:Gtt}
\end{equation}
Insert Eq. (\ref{eq:Gtt}) to Eq. (\ref{eq:ntt}), we have
\begin{equation}
    \langle n_{i\sigma}\rangle=-i\int\frac{d\omega}{2\pi}G_{i\sigma}^{<}(\omega).
\end{equation}
Use the thermal equilibrium relations $G_{i\sigma}^{<}(\omega)=f(\omega)[G_{i\sigma}^a(\omega)-G_{i\sigma}^r(\omega)]$ and $G_{i\sigma}^a(\omega)=[G_{i\sigma}^r(\omega)]^{*}$, we arrive at
\begin{equation}
    \langle n_{i\sigma}\rangle=-\frac{1}{\pi}\int d\omega f(\omega)\text{Im}[G_{i\sigma}^r(\omega)].
\end{equation}\\

Similarly, for the general operator $D=n_{1\sigma_1}n_{2\sigma_2}...n_{i_N\sigma_N}$ with $i_N\sigma_N\neq i\sigma$, we have the set of Green functions
\begin{align}
    G_{Dd}^{<}(t,t^{'})&=i\langle d_{i\sigma}^\dag(t^{'})D(t)d_{i\sigma}(t)\rangle,\\
    G_{Dd}^{r}(t,t^{'})&=-i\Theta(t-t^{'})\langle \{D(t)d_{i\sigma}(t),d_{i\sigma}^\dag(t^{'})\}\rangle,\label{eq:Gr}\\
    G_{Dd}^{a}(t,t^{'})&=i\Theta(t^{'}-t)\langle\{D(t)d_{i\sigma}(t),d_{i\sigma}^\dag(t^{'})\}\rangle.\label{eq:Ga}
\end{align}
Apply $t=t^{'}$, the lesser Green function becomes
\begin{align}
    G_{Dd}^{<}(t,t)&=i\langle d_{i\sigma}^\dag(t)D(t)d_{i\sigma}(t)\rangle,\\
    \langle D(t)n_{i\sigma}(t)\rangle&=\langle D d_{i\sigma}^\dag d_{i\sigma}\rangle=-iG_{Dd}^{<}(t,t).\label{eq:Dna}
\end{align}

Within the same density-sector closure used in the EOM hierarchy,
$[\hat D,\hat H]\simeq 0$, the Heisenberg equation
$\partial_t\hat D(t)=(i/\hbar)[\hat H,\hat D(t)]$ gives
$\partial_t\hat D(t)\simeq0$. Thus, within the same approximation,
the density product is treated as time independent,
$\hat D(t)\simeq\hat D$. Together with
$\hat D^\dagger=\hat D$ and
$[\hat D,\hat d_{i\sigma}]
=[\hat D,\hat d^\dagger_{i\sigma}]=0$, this allows us to establish
the relation between the generalized retarded and advanced Green
functions. Setting $t'=0$ in Eq. (\ref{eq:Gr}) and Eq. (\ref{eq:Ga}), we have
\begin{equation}
    G_{Dd}^{r}(t)=-i\Theta(t)\langle \{D(t)d_{i\sigma}(t),d_{i\sigma}^\dag(0)\}\rangle.
\end{equation}
\begin{equation}
    G_{Dd}^{a}(t)=i\Theta(-t)\langle\{D(t)d_{i\sigma}(t),d_{i\sigma}^\dag(0)\}\rangle
\end{equation}
Take the complex
conjugate and use $[\hat{D},\hat{d}_{i\sigma}]=0$, we obtain
\begin{align}
\left[G^r_{Dd}(t)\right]^*
&=
i\Theta(t)
\langle
\{
\hat d_{i\sigma}(0),
\hat d^\dagger_{i\sigma}(t)\hat D
\}\rangle
\nonumber\\
&=
i\Theta(t)
\langle
\{
\hat d_{i\sigma}(0),
\hat D\hat d^\dagger_{i\sigma}(t)
\}\rangle .
\end{align}
Applying equilibrium time-translation invariance to get $\langle \{ \hat d_{i\sigma}(0), \hat D \hat d^\dagger_{i\sigma}(t) \} \rangle = \langle \{ \hat d_{i\sigma}(-t), \hat D \hat d^\dagger_{i\sigma}(0) \} \rangle$ and
using $\hat D(t)\simeq\hat D$, this becomes
\begin{align}
\left[G^r_{Dd}(t)\right]^*
&=
i\Theta(t)
\langle\{
\hat d_{i\sigma}(-t),
\hat D\hat d^\dagger_{i\sigma}(0)\}\rangle
\nonumber\\
&=
i\Theta(t)
\langle
\{
\hat D\hat d_{i\sigma}(-t),
\hat d^\dagger_{i\sigma}(0)
\}\rangle
=
G^a_{Dd}(-t).
\end{align}
Consequently, Fourier transformation gives
\begin{align}
\left[G^r_{Dd}(\omega)\right]^*
&=
\int_{-\infty}^{+\infty} dt\,
e^{-i\omega t}
\left[G^r_{Dd}(t)\right]^*
\nonumber\\
&=
\int_{-\infty}^{+\infty} dt\,
e^{-i\omega t}
G^a_{Dd}(-t)
=
G^a_{Dd}(\omega),
\end{align}
where the last equality follows by $t\rightarrow -t$. {Performing} the Fourier transform of Eq. (\ref{eq:Dna}) and use $G_{Dd}^{<}(\omega)=f(\omega)[G_{Dd}^a(\omega)-G_{Dd}^r(\omega)]$ and $G_{Dd}^a(\omega)=[G_{Dd}^r(\omega)]^{*}$, we arrive at 
\begin{equation}
    \langle Dn_{i\sigma}\rangle=-\frac{1}{\pi}\int d\omega f(\omega)\text{Im}[G_{Dd}^r(\omega)].
\end{equation}

\end{widetext}
\end{document}